\documentclass[reprint,pra,aps,longbibliography,showpacs]{revtex4-1} 
\usepackage{silence}
\usepackage{amsmath,revsymb,graphicx,amssymb,array,natbib}

\newcommand{\kt}[1]{\ensuremath{|#1\rangle}}
\newcommand{\br}[1]{\ensuremath{\langle#1|}}
\newcommand{\bk}[2]{\ensuremath{\langle #1|#2\rangle}}
\newcommand{\HS}{\mathcal{H}}

\begin{document}

\title{Infinite Barriers and Symmetries for a\\ Few Trapped Particles in One Dimension}

\author{N.L.~Harshman\email{Electronic address: harshman@american.edu}}

\affiliation{Department of Physics, 
4400 Massachusetts Ave.\ NW, American University, Washington, DC 20016-8058}

\begin{abstract}
This article investigates the properties of a few interacting particles trapped in a few wells and how these properties change under adiabatic tuning of interaction strength and inter-well tunneling. While some system properties are dependent on the specific shapes of the traps and the interactions, this article applies symmetry analysis to identify generic  features in the spectrum of stationary states of few-particle, few-well systems. Extended attention is given to a simple but flexible three-parameter model of two particles in two wells in one dimension. A key insight is that two limiting cases, hard-core repulsion and no inter-well tunneling, can both be treated as emergent symmetries of the few-particle Hamiltonian. These symmetries are the mathematical consequences of infinite barriers in configuration space. They are necessary to explain the pattern of degeneracies in the energy spectrum, to understand how degeneracies are broken for models away from limiting cases, and to explain separability and integrability. These symmetry methods are extendable to more complicated models and the results have practical consequences for stable state control in few-particle, few-well systems with ultracold atoms in optical traps.

\end{abstract}

\maketitle

\section{Introduction}

One motivation for the symmetry analysis of few-body, few-well models is experiments with a few ultracold atoms trapped in an optical potential, either a single well or an arrangement of wells like a lattice or crystal. In some recent experiments~\cite{serwane_deterministic_2011, wenz_few_2013, kaufman_two-particle_2014, kaufman_two-particle_2014, murmann_two_2015, murmann_antiferromagnetic_2015, kaufman_entangling_2015}, a deterministic number of atoms are loaded into a trap, and then the shape and arrangement of the wells and the strength of the interparticle interaction are changed.

These experiments, with the already-exquisite and always-improving control they offer, present an irresistible playground for pure and applied quantum theorists for at least three reasons:
\begin{itemize}
\item Starting with a few particles and a few wells, we can take a `bottom-up' approach to studying many-body physics and emergent phenomena. Models with strong interactions and multiple competing length scales and energy scales can be difficult to characterize and solve, especially identical particles with internal degrees of freedom. However, these kinds of models are important in condensed matter physics and their dynamical and thermodynamical properties are rich and varied. For example, few-body and effectively one-dimensional single-well and few-well systems have already been used to investigate magnetism and quantum phase transitions in Heisenberg spin chain models~\cite{bugnion_ferromagnetic_2013-1, cui_ground-state_2014, volosniev_strongly_2014, deuretzbacher_quantum_2014, massignan_magnetism_2015, grining_crossover_2015, beverland_realizing_2016} and to check the consistency of the approximations used to solve Hubbard-type models \cite{garcia-march_macroscopic_2012, gillet_tunneling_2014, Santos2015186, sowinski_diffusion_2016}. 
\item The control possible over a few atoms in a few wells allows unprecedented possibilities for quantum state preparation and manipulation. For example, there are protocols to construct highly-entangled NOON states from a few particles in a double-well \cite{garcia-march_macroscopic_2012, PhysRevA.92.033621}. This kind of multiparticle coherence has been demonstrated to be a resource for application in quantum sensing and measurement and for other quantum information processing tasks~\cite{divincenzo_universal_2000, PhysRevLett.91.207901, PhysRevA.91.023620}. This article provides an specific example of a quantum state control mechanism that generates superpositions using cyclic adiabatic tuning. Also, a speculative proposal explored below uses the large but controllable degrees of freedom of a few-well, few-body system to embody combinatoric problems in a quantum systems.
\item Finally, generalizing the previous two reasons, systems of ultracold atoms in tunable traps with tunable interactions allow us to explore and test quantum dynamics like never before. The consequences of integrability and chaos, the interrelations among interaction, indistinguishability, and entanglement, and other questions at the boundaries of quantum mechanics can be directly interrogated in the cold-atom laboratory.
\end{itemize}
As the number of particles and wells increase, the degrees of freedom and complexity of the problem grow exponentially. This makes analytical and even numerical progress difficult for the bottom-up approach except in certain limiting cases with enhanced solvability, such as contact interactions in harmonic traps or infinite square well. This motivates the need for generating solvable few-atom, few-well models, and for identifying universal features on non-solvable models. Symmetry analysis is key to both of these tasks.

The primary focus of this article is one-dimensional, double-well models. As a motivating example, consider the experiment described in Ref.~\cite{murmann_two_2015}: two fermions are loaded into an effectively one-dimensional a double well with tunable shape and interaction strength. Even though such systems have just two degrees of freedom, they have been a testing ground for quantum dynamics since the beginning of the subject. Some recent relevant investigations on the dynamics of two (or a few) particles in a double well include analyses of interaction-influenced tunneling~\cite{liu_two_2015, sowinski_diffusion_2016}, spatial state superpositions~\cite{garcia-march_macroscopic_2012}, quantum integrability~\cite{braak_integrability_2014, cosme_relaxation_2015, ymai_quantum_2016}, state control and logic gates~\cite{hayes_quantum_2007}, phase measurements~\cite{das_unitary_2013}, interference dynamics~\cite{kaufman_two-particle_2014, mullin_interference_2015}, fermionic quantum number pinning~\cite{schilling_hubbard_2015} and entanglement and quantum information theory~\cite{sowinski_dynamics_2010, brandt_double-well_2015, kaufman_entangling_2015, schilling_hubbard_2015, lingua_delocalization_2016}. Some of these references are motivated by applications to many-atom gases in arrays of double-well optical traps (for example, see \cite{albiez_direct_2005, anderlini_controlled_2007, trotzky_time-resolved_2008}). One-dimensional double-well models also have applications as effective theories for systems with more degrees of freedom (e.g.\ ion chains in Paul traps~\cite{retzker_double_2008}, Josephson junctions, spontaneous symmetry breaking, etc.) and are extensively used as pedagogical examples~\cite{foot_double_2011, jelic_double-well_2012}.

This article attempts to organize this double-well phenomenology using symmetry methods. Particle exchange and parity are familiar symmetries and certainly useful, but this article goes beyond these and describes how the special cases of (1) no interactions, (2) no tunneling, and (3) hard core interactions can be formulated as kinematic symmetries of the Hamiltonian. These symmetries are used to characterize the spectrum of double-well, interacting particle Hamiltonians and to make spectral maps among models related by symmetry-breaking. This analysis reveals what dynamical effects are particular to specific trap or interaction, and what are generic. These symmetries can be exploited to enhance analytical methods like exact diagonalization, perturbation theory and variational methods. Finally, this article shows how the symmetries that are preserved under parametric variation between two models can be exploited for adiabatic state control.

\subsection{The Model}

\begin{figure}
\includegraphics[width=\columnwidth]{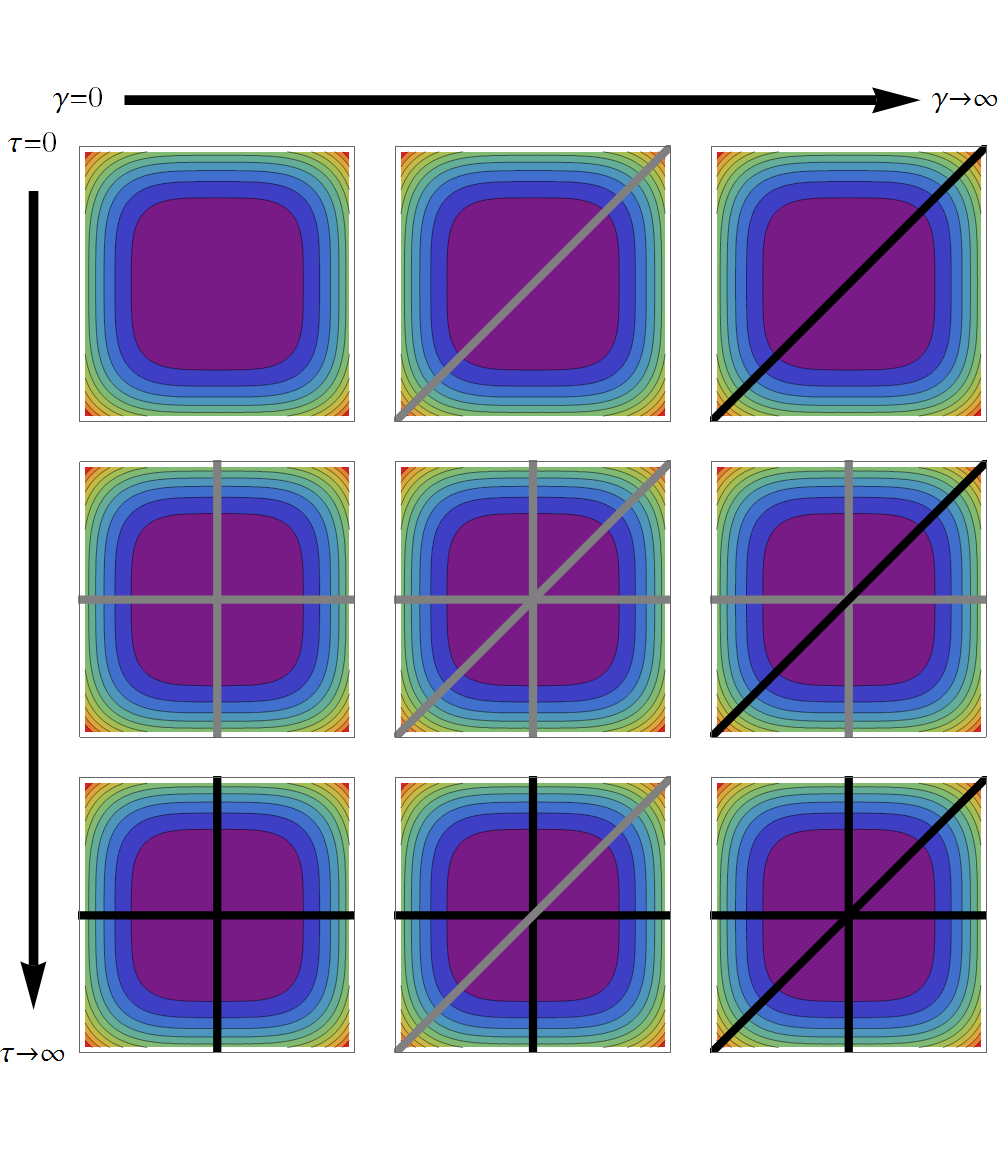}%
\caption{The potential energy of Hamiltonian %(\ref{eq:habg})
  for a quartic trap with $a=0$. Black lines represent impenetrable barriers; gray lines are finite barriers that allow tunneling. The first column depicts $H^0_0$, $H^\tau_0$, and $H^\infty_0$; these three Hamiltonians are integrable for any trap shape and for any number of identical particles. The last column depicts the unitary limit of contact interactions $H^0_\infty$, $H^\tau_\infty$, and $H^\infty_\infty$; these three Hamiltonians are also integrable for any trap shape or number of particles. The middle column with arbitrary interaction strength are $H^0_\gamma$, $H^\tau_\gamma$, and $H^\infty_\gamma$. Generally, these Hamiltonians are not integrable, but $H^0_\gamma$ and $H^\infty_\gamma$ are integrable for two particles in a harmonic trap or any number of particles in the infinite square well trap. \label{fig:quartic}}
\end{figure}

This article considers a three-parameter model for two interacting particles in two wells:
\begin{eqnarray}\label{eq:habg}
H^\tau_\gamma &= & - \frac{\hbar^2}{2 m} \left( \frac{\partial^2}{\partial x_1^2} + \frac{\partial^2}{\partial x_2^2}  \right) + V(x_1) + V(x_2)\nonumber\\
&&{} + \tau \left( \delta(x_1 - a) + \delta(x_2 -a ) \right) + \gamma\delta(x_1 - x_2).
\end{eqnarray}
This Hamiltonian describes a symmetric trap $V(x) = V(-x)$ that is split by a delta-function barrier with strength $\tau$ a distance $a$ from the middle (see \cite{belloni_infinite_2014, glasser_energy_2015} for general analysis of traps split by delta-barriers). The two particles experience a contact interaction with strength $\gamma$. Fig.~\ref{fig:quartic} schematically depicts the potential energy of Hamiltonian (\ref{eq:habg}) as a contour plot in configuration space for a purely quartic trap when $a=0$. The corners of Fig.~\ref{fig:quartic} represent the four limiting cases of $H^\tau_\gamma$: no interactions and no barrier $H^0_0$; no interactions and no tunneling (infinite barriers) $H^\infty_0$; unitary limit of contact interactions and no barrier $H^0_\infty$; and unitary interactions and no tunneling $H^\infty_\infty$. Tab.\ \ref{tab:deg} summarizes the degeneracies in the energy spectrum of some of these limiting cases of (\ref{eq:habg}). 

In principle, how spectral properties change as the Hamiltonian $H^\tau_\gamma$ is tuned can be inferred from symmetry. For the Hamiltonian (\ref{eq:habg}) many systematic degeneracies (as opposed to accidental degeneracies, a distinction clarified below) are independent of the specific trap shape $V(x)$; see Tab.\ \ref{tab:deg}. These degeneracies can be explained using the extra symmetries that arise from impenetrable barriers. Further, how these levels map to one another under adiabatic changes of parameters sometimes can be established from symmetry alone. Other times, such energy level mappings must be completed with the assistance of perturbation theory, exact diagonalization, variational methods, or other numerical approximation schemes, all of which are simplified using symmetry. Adiabatic mappings can then provide pathways for coherent state control, especially when combined with spin or other internal states that affect the allowed state space of identical particles through bosonic or fermionic symmetrization. In full disclosure, many of these tasks can be efficiently accomplished numerically without exploiting all the symmetries of a Hamiltonian for two particles in two wells. However, as we build up from the bottom, using group theory to turn combinatorics into algebraic observables on the Hilbert space may allow us to push deeper into the emergence of many-body phenomena.

\begin{table}%
\begin{tabular}{cccc}
\hline
Hamiltonians & Parameters & \multicolumn{2}{c}{Degeneracies} \\
&& $a \neq 0$ & $a=0$ \\
\hline
$H^0_0$, $H^\tau_0$ & $\tau \geq 0$, $\gamma=0$ & $1,2$ & $1,2$ \\
$H^0_\gamma$, $H^\tau_\gamma$ & $\tau \geq 0$, $\gamma\neq 0$ & $1$ & $1$ \\
$H^0_\infty$, $H^\tau_\infty$ & $\tau \geq 0$, $\gamma\to \infty$ & $2$ & $2$ \\
$H^\infty_0$ & $\tau \to \infty$, $\gamma = 0$ & $1,2$ & $4,8$ \\
$H^\infty_\gamma$ & $\tau \to \infty$, $\gamma \neq 0$ & $1,2$ & $2,6$ \\
$H^\infty_\infty$ & $\tau \to \infty$, $\gamma \to \infty$ & $2$ & $2,8$ \\
\end{tabular}
\caption{Systematic degeneracies of Hamiltonian %(\ref{eq:habg})
for various values of the parameters of the barrier location $a$, zero-range barrier strength $\tau$, and zero-range interaction strength $\gamma$. }
\label{tab:deg}
\end{table}

\subsection{Kinematic Symmetries}

The kinematic symmetry group of the Hamiltonian (i.e.\ the set of all unitary operators that map all energy eigenstates into energy eigenstates with the same energy) can be used to classify states, categorize types of degeneracies, and select useful observables~\footnote{For contrast, the dynamic symmetry group of the Hamiltonian includes operators that map all energy eigenstates into energy eigenstates with not necessarily the same energy. This group can be used to generate the spectrum of the Hamiltonian from a smaller set of states.}. In principle, if the correct kinematic symmetry is found, then every degenerate energy subspace of the will carry an irreducible representation (irrep) of this symmetry. This article describes and exploits the following three kinematic symmetries:
\begin{itemize}
\item The \emph{symmetry of separability}: Identical non-interacting particles have independent time evolutions. In particular, when the Hamiltonian splits into a sum of identical sub-Hamiltonians, then this symmetry can be described by the wreath product $\mathrm{T}_t \wr \mathrm{S}_N $, where $\mathrm{S}_N$ is the symmetric group on $N$ particles, $\mathrm{T}_t$ is the time-translation subgroup for each independent particle, and $\wr$ is the wreath product that interweaves $\mathrm{S}_N$ with $N$ copies of $\mathrm{T}_t$ (described in more detail below). Interactions break this symmetry into the subgroup $\mathrm{T}_t \times \mathrm{S}_N$~\cite{ Harshman2016_I, Harshman2016_II}.
\item \emph{Well permutation symmetry}: When there is no tunneling, then each well has independent dynamics. If there are $M$ identical wells, then there is a symmetry isomorphic to $\mathrm{S}_M \wr \mathrm{T}_t$ that corresponds to permuting the the individual, disconnected wells. Tunneling breaks this local symmetry in a manner that depends on the global structure of the wells~\cite{Harshman2017}.
\item \emph{Ordering permutation symmetry}: $N$ particles in a one-dimensional single well with hard-core interactions cannot tunnel past each other. Therefore, there are $N!$ identical $N$-dimensional wells corresponding to each dynamically-stable possible ordering and the symmetry has the form $\mathrm{T}_t \times \mathrm{S}_{N!}$~\cite{Harshman2017}. In the near unitary limit, this symmetry is broken in a way that can be calculated exactly for contact interactions~\cite{volosniev_strongly_2014, deuretzbacher_quantum_2014, Harshman2016_II}.
\end{itemize}

The second and third of these symmetries depend on the presence of infinite barriers in space or configuration space. To understand  the importance of infinite barriers for symmetry in few-atom, few-well system, consider the following one-particle, one-dimensional harmonic Hamiltonian:
\begin{equation}\label{eq:ham1}
H = -\frac{1}{2m} \frac{\partial^2}{\partial x^2} + \frac{1}{2}m\omega^2 x^2 + \tau \delta(x).
\end{equation}
Analytic solutions for the eigensystem of this Hamiltonian are calculable for any value of $\tau$ to arbitrary accuracy by solving a transcendental equation for energy~\cite{avakian_spectroscopy_1987, busch_two_1998}. However, in the limit $\tau \to \infty$, the harmonic trap splits into two identical wells with no tunneling allowed. Each side of the barrier is dynamically decoupled from the other and their relative phases have no physical meaning; i.e.~the two sides cannot compare clocks. The stationary states when $\tau \to \infty$ can be found by patching together the odd solutions of (\ref{eq:ham1}) with $\tau=0$ into two solutions (one with even parity, one with odd). Every energy level in the spectrum is doubly-degenerate. The singularity at $x=0$ severs the configuration space by forcing a node onto every wave function with finite energy, and this invalidates the standard proof that all bound-state energy eigenfunctions in one-dimension are non-degenerate~\cite{loudon_one-dimensional_1959}.

Degeneracies must be consistent with the dimensions of irreps kinematic symmetry group of the Hamiltonian. In the case $\tau \to \infty$ the double-degeneracy can be understood as arising from the dynamical decoupling of the two domains $x>0$ and $x<0$. In each domain, time evolution by an amount $t$ is represented by a unitary operator $U_{x>0}(t)$ or $U_{x<0}(t)$ that rotates the phase of the wave function only on one side of the barrier. These two operators commute with the total Hamiltonian and with each other. This doubling of time evolution symmetry can explain the doubling of the spectrum.

Well permutation symmetry manifests whenever identical traps are dynamically decoupled, and it occurs in any dimension. In the limit of impenetrable one-dimensional particles, the Hamiltonian similarly decouples into independent subsystems. The few-particle configuration space is sectioned into fixed orderings by the interactions, and these sections are equivalent and exchangeable for identical particles. Each ordering is a `subsystem' that is decoupled from all the other orderings. If the particles are totally or partially indistinguishable, then there are phase relations among different orderings that are induced by particle exchange symmetries. In the case of totally indistinguishable bosons and fermions, these phase relations lift the degeneracy completely (i.e.~the famous Bose-Fermi mapping of Girardeau~\cite{girardeau_relationship_1960}), otherwise it is more complicated for more than two particles~\cite{deuretzbacher_exact_2008, fang_exact_2011, Harshman2016_I, Harshman2016_II}.

\subsection{Outline of Paper}

The outline of this paper is as follows. Sect.~\ref{sect:onepart} goes `beneath the bottom' and looks at symmetries for one particle in double-well scenarios. Besides the familiar global symmetries of parity and particle exchange, infinite barriers induce piecewise-linear \emph{local} symmetries. Local symmetries have been shown to be useful for wave propagation and scattering in one-dimensional systems with partial symmetry~\cite{kalozoumis_invariants_2014-1, kalozoumis_invariant_2015}. This section also clarifies how this article makes the distinction between systematic and accidental symmetries (following the definitions in \cite{shaw_degeneracy_1974}). Sect.~\ref{sect:twopart} describes symmetries and symmetry breaking for two interacting particles in an arbitrary double-well system. Both Sect.~\ref{sect:onepart} and \ref{sect:twopart} focus on the puzzle of identifying sufficient kinematic symmetries to explain spectral degeneracies.  Readers less interested in the structural analysis of symmetries and irreps may want to skim these sections and start in Sect.~\ref{sect:habg}, which applies these symmetries to analyze the three parameter model (\ref{eq:habg}). For this model, the zero-range nature of the barrier and interaction provides additional symmetry, leading to integrability and solvability for a variety of limiting cases. In Sect.~\ref{sect:solv}, the additional symmetry provided when the trap potential $V(x)$ is the infinite square well or the harmonic trap is exploited to construct explicit adiabatic maps between limiting cases. Sect.~\ref{sect:multi} briefly indicates how these ideas can be extended to a few particles in a few wells, and the concluding section provides an outlook on possible extensions and applications of this work.

\section{Symmetries for one particle in two wells}\label{sect:onepart}

\begin{figure}%
\includegraphics[width= \columnwidth]{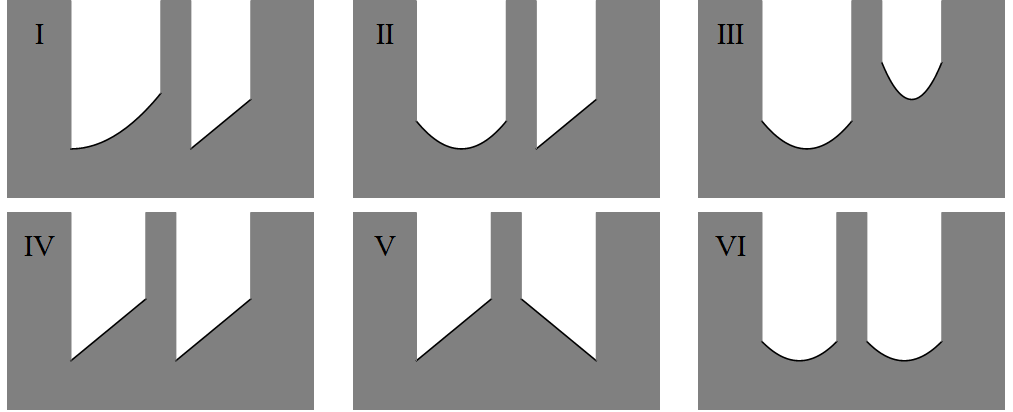}%
\caption{Schematic representation of six symmetry cases of double wells in one dimension with different configuration space and kinematic symmetries, described in the text.}%
\label{fig:6types}%
\end{figure}

From the perspective of kinematic symmetries, there are six kinds of impenetrable double wells in one-dimension. They are distinguished depending on whether the wells have the same shape, whether the wells are individually symmetric under reflection, and whether the pair of wells is symmetric under a global reflection. The six distinct possibilities are depicted schematically in Fig.\ \ref{fig:6types}. In the top three, the two wells are different and have different energy spectra. For the bottom three, wells are identical (or mirror images, as in case V).

\subsection{Configuration space symmetries}

Consider the \emph{configuration space symmetry group} for one particle in a double well. Configuration space symmetries are transformations of one-particle configuration space $\mathcal{X}\sim\mathbb{R}$ that commute with the Hamiltonian and are a subgroup of the kinematic symmetry. Equivalently, configuration space symmetry transformations map stationary state wave functions into other stationary state wave functions with the same energy. Following~\cite{kalozoumis_invariants_2014-1}, we further distinguish \emph{global} and \emph{local} configuration space symmetries. An example of a global symmetry transformation is a linear transformation of the entire space $\mathcal{X}$  that commutes with the Hamiltonian. Generally in one dimension, any global linear transformation is a translation, reflection, or glide reflection. For double wells, only the global reflection is possible. Local symmetry transformations include \emph{piecewise} linear transformations. For example, a transformation like ``apply a reflection to well $a$ but leave well $b$ alone''. Local symmetries are only possible when there are infinite barriers and the configuration space wave functions have nodes at the boundaries between the different domains of the piecewise transformations. The one-particle configuration space $\mathcal{X}$ can be divided into three domains: the left well $\mathcal{X}_a$, the barrier $\mathcal{X}_\tau$, and the right well $\mathcal{X}_b$.

Using these definitions, consider the six cases, also summarized in Tab.~\ref{tab:1partsym}:
\begin{itemize}
\item Case I: There is no global or local symmetry, so the configuration space symmetry group is the trivial group $\mathrm{E}$ of just the identity transformation on $\mathcal{X}$, whether or not there is tunneling. 
\item Case II: There is a single local symmetry transformation, a parity reflection $\pi_a$ in domain $\mathcal{X}_a$. The configuration space symmetry is denoted $\mathrm{O}(1)_a$ and has order 2. When tunneling occurs, this local symmetry is broken.
\item Case III: Each well is parity symmetric, so without tunneling there are two local reflections $\pi_a$ and $\pi_b$, giving $\mathrm{O}(1)_a\times \mathrm{O}(1)_b \sim \mathrm{Z}_4$, where $\mathrm{Z}_4$ is the abstract cyclic group of order 4. Again, tunneling breaks the local symmetry and there are no global configuration space symmetries.
\item Case IV: There are no single-domain local symmetries, but domain $\mathcal{X}_a$ can be translated right and domain $\mathcal{X}_b$ can be translated left. Call this piecewise linear transformation $w_{ab}$ and call the order-2 group it generates $\mathrm{W}_2 \sim \mathrm{Z}_2$. If there is tunneling, then this discontinuous transformation will not map stationary states into stationary states, and there is no global symmetry.
\item Case V: The domains can be flipped and then exchanged, i.e. the transformation $\pi_a\pi_b w_{ab}$. This is the same abstract group as case IV, but denote it by $\mathrm{W}_2'$ to signify that it is a different realization. The transformation $\pi_a\pi_b w_{ab}$ is equivalent to the total parity transformation $\pi$. This global symmetry transformation is preserved in the presence of tunneling.
\item Case VI: Counting the identity $e$, there are eight configuration space symmetry transformations for impenetrable wells: $\pi_a$, $\pi_b$, $\pi_a\pi_b$, $w_{ab}$, $\pi_a w_{ab}$, $\pi_b w_ab$, $\pi_a \pi_b w_{ab}$. This group is isomorphic to $\mathrm{D}_4$, the dihedral group with four reflections, also realized as the two-dimensional point symmetries of a square. Unlike the other five cases, this group is not abelian, for example $\pi_a w_{ab} = w_{ab} \pi_b$. As with case V, tunneling breaks almost all of these symmetries, leaving only global parity.
\end{itemize}

\begin{table}%
\begin{tabular}{c|c|c|c|c}
\hline
Type & $\mathrm{C}_1^\infty$ & $|\mathrm{C}_1^\infty|$ & $\mathrm{C}_1^\tau$ & $|\mathrm{C}_1^\tau|$ \\
\hline
I & $\mathrm{E}$ & $1$ & $\mathrm{E}$ & $1$ \\
II & $\mathrm{O}(1)_a$ & $2$ & $\mathrm{E}$ & $1$  \\
III & $\mathrm{O}(1)_a\times \mathrm{O}(1)_b$ & $4$ & $\mathrm{E}$ & $1$  \\
IV & $\mathrm{W}_2$ & $2$ & $\mathrm{E}$ & $1$  \\
V & $\mathrm{W}'_2$ & $2$ & $\mathrm{O}(1)$ & $2$ \\
VI & $\mathrm{O}(1)\wr\mathrm{W}_2$ & $8$ & $\mathrm{O}(1)$ & $2$ \\
\hline
\end{tabular}
\caption{This table lists the configuration space symmetry for one particle in two well for the case of no tunneling $\mathrm{C}_1^\infty$ and tunneling $\mathrm{C}_1^\tau$. The groups are expressed using notation explained in the text. The order of the groups (i.e. number of elements including the identity) is listed in the column after each group. In each case, $\mathrm{C}_1^\tau$ is the subgroup of $\mathrm{C}_1^\infty$ preserved when the no-tunneling symmetry is broken.}
\label{tab:1partsym}
\end{table}

One way to describe the symmetry of case VI is using the wreath product $\wr$, a kind of semidirect product that is used in combinatorics to describe the permutations of objects with structure. In case VI, the configuration space symmetry group can be expressed as $\mathrm{O}(1)\wr \mathrm{W}_2$. The first term in the wreath product is the symmetry of the well; the second term is the exchange symmetry of the wells. Generally, in the wreath product   $\mathrm{G} \wr \mathrm{S}_N$ between a finite group $\mathrm{G}$ with order $g$ and a permutation group $\mathrm{S}_N$ with order $N!$, the order of the wreath product is $(g^N \cdot N!)$. A well-known example of a wreath product is the hyperoctahedral group $\mathrm{Z}_2 \wr \mathrm{S}_N$, the symmetry of an $N$-dimensional cube. Other examples of wreath products include the lamplighter group $\mathrm{Z}_2 \wr \mathrm{Z}$ and the group of transformations of a Rubik's cube~\cite{joyner_adventures_2008}. In the next section, the wreath product structure is used to generalize local symmetries to the few-body case.

\subsection{Degeneracies and kinematic symmetries}

In the limit of no tunneling, each well is assumed have a  discrete and singly-degenerate spectrum: $\sigma_a = \{\alpha_0, \alpha_1, \alpha_2, \ldots \}$ for the left well and $\sigma_b = \{\beta_0, \beta_1,  \ldots \}$ for the right well. For cases IV, V, and VI, the spectra  are the same $\sigma_a = \sigma_b$. When wells are not the same shape, they may have energy levels that line up `on accident', but in general this requires highly specific fine-tuning. For example, the depth, width, or other shape parameters are tuned `just right' so that one or more levels line up perfectly. Counterexamples of different-shaped wells with many overlapping levels are wells that are supersymmetric pairs, or infinite square wells with rational ratios, but these seemingly `accidental' degeneracies derive from dynamical or spectrum-generating symmetries, and are in fact systematic.

So when there is no tunneling, and barring accidental degeneracies, the double-well energy spectra $\sigma_1 = \sigma_a \cup \sigma_b$ for cases I-III are singly-degenerate and for cases IV-VI the spectra $\sigma_1 = \sigma_a = \sigma_b$ is doubly-degenerate. Systematic degeneracies like these should be explained by the kinematic symmetry group of the Hamiltonian. When the correct kinematic symmetry group has been identified, the degeneracies of the energy levels correspond to the dimensions of the unitary irreducible representations (irreps) of the symmetry group~\footnote{Note that there can be irreps of the symmetry group that do not correspond to physically realizable states.}. Configuration space symmetries alone cannot explain the two-fold degeneracy for cases IV-VI. The configuration space symmetry groups for cases IV and V only have one-dimensional irreps. The configuration space symmetry group for case VI is isomorphic to the symmetry of a square and does have a two-dimensional irrep, but explicit construction of wave functions shows that not all double-degenerate energy levels correspond to that irrep~\cite{Harshman2016_I}.

So what symmetry group gives the correct two-fold systematic degeneracies for cases IV-VI when there is no tunneling? As mentioned in the introduction, the solution to this puzzle lies in the observation that in the absence of tunneling the two wells are \emph{dynamically independent}. The one-particle Hamiltonian $\hat{h}$ can be decomposed into a sum of sub-Hamiltonians $\hat{h}_a$ and $\hat{h}_b$ by restricting the position representation~\footnote{A note about notation: an operator with a hat $\hat{O}$ is the operator realized as a linear transformation of the system Hilbert space. The same operator without a hat $O$ is the position-space representation of that operator.}
of the Hamiltonian $h(x)$
\begin{equation}
\br{x} \hat{h} \kt{x'} = h(x) \delta(x-x')
\end{equation} 
to the well regions $\mathcal{X}_a$ and $\mathcal{X}_b$:
\begin{eqnarray}
h_a(x) &=& h(x)|_{\mathcal{X}_a} \Rightarrow \hat{h}_a\nonumber\\
h_b(x) &=& h(x)|_{\mathcal{X}_b} \Rightarrow \hat{h}_b.
\end{eqnarray}
 These sub-Hamiltonians are defined on disjoint domains and they commute with each other and the total Hamiltonian. They each generate a time translation operator $\hat{U}_a(t) = \exp(-i \hat{h}_a t/\hbar)$. As unitary operators that commute with $\hat{h}$, they generate a subgroup of the kinematic symmetry group. For case I (which has the least symmetry), the kinematic symmetry group is the product of each well's time translation group $\mathrm{T}_a \times \mathrm{T}_b$~\footnote{The notation $\mathrm{T}_a$ for the time translation group for well $a$ and $\mathrm{T}_b$ for well $b$ is a bit misleading. Both groups are isomorphic to the translation group by a single real parameter. Their difference is in the possible characters of the irreps, where the characters are just the allowed energies in the spectra $\sigma_a$ and $\sigma_b$. So the notation like $\mathrm{T}_a$ is useful for bookkeeping, but it does mix together the group and the irrep notation in a way that is a little fishy.}. For cases II and III, the kinematic symmetry groups for no-tunneling are $\mathrm{O}(1)_a \times \mathrm{T}_a \times \mathrm{T}_b$ and $\mathrm{O}(1)_a \times \mathrm{O}(1)_b \times \mathrm{T}_a \times \mathrm{T}_b$, respectively. In all three cases, these groups only have irreps that are one-dimensional, and spectra that are therefore non-degenerate except for `accidents'.

For cases IV and V, the time translation group irreps are the same $\mathrm{T}_a = \mathrm{T}_b$ and the kinematic symmetry group can be expressed as $\mathrm{T}_a\wr \mathrm{W}_2$ and $\mathrm{T}_a\wr \mathrm{W}'_2$, respectively. Case VI is $(\mathrm{T}_a\times \mathrm{O}(1) )\wr \mathrm{W}_2$. These groups have two-dimensional irreps that correspond to the familiar energy eigenspaces spanned by left-right localized basis states~\cite{PhysRevA.92.033621}.

When tunneling is present, coupling of the wells breaks local time-translation symmetry. The kinematic symmetry for cases I-IV is just time translation $\mathrm{T}$ and for cases V-VI it is global time translation and parity $\mathrm{T}\times \mathrm{O}(1)$. These groups only have one-dimensional irreps, and therefore tunneling splits the the doubly-degenerate levels for cases IV-VI. The splitting energy cannot be inferred from symmetry, although the familiar result that the parity symmetric states for cases V and VI have lower energy can be derived from symmetry alone. 

\section{Symmetries for two particles in two wells}\label{sect:twopart}

\begin{figure}
\includegraphics[width= \columnwidth]{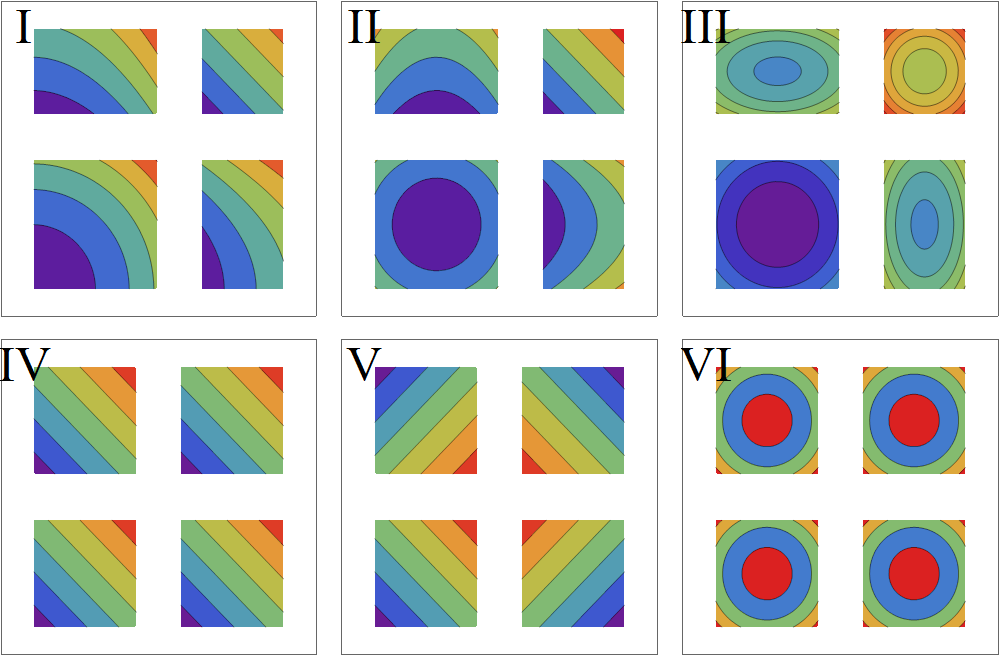}%
\caption{Contour plots of the potentials for two particles in each of the six double-well traps types in Fig.~\protect\ref{fig:quartic}
 without interactions. In each subfigure, sector in the upper right corner  is the domain $\mathcal{X}_A$, where both particle are in the right well $b$, and the upper right corner is domain $\mathcal{X}_C$, both particles in left well $a$. The off-diagonal domains $\mathcal{X}_B$ and $\mathcal{X}_D$ correspond to the particles in different wells. Note that without interactions, all four sectors are separable in all six cases.}
\label{fig:types2}%
\end{figure}

\begin{figure}
\includegraphics[width= \columnwidth]{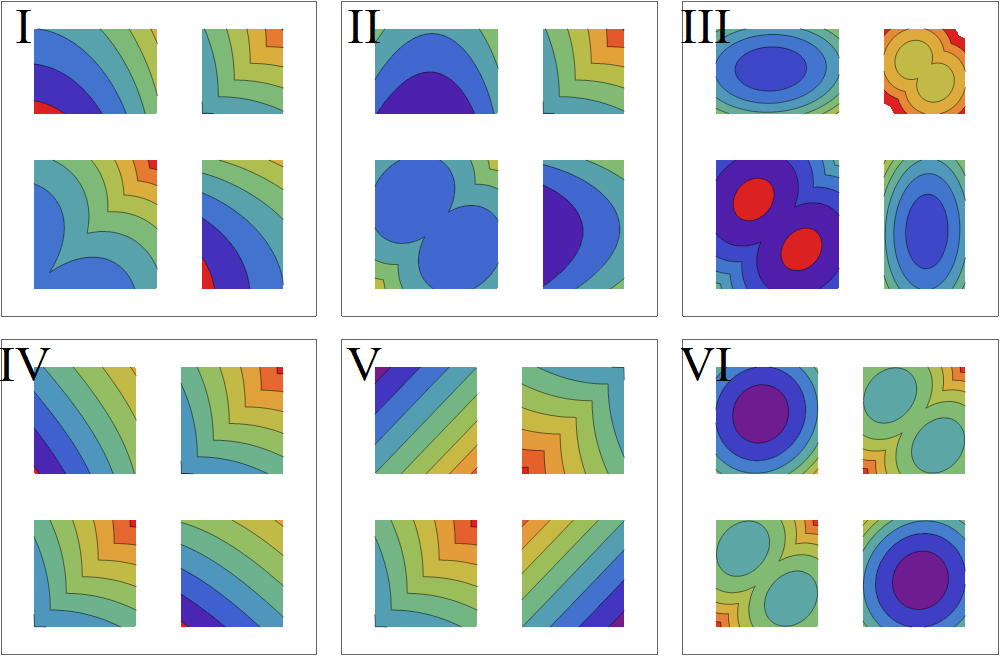}%
\caption{Contour plots of the potentials for two particles in each of the six double-well traps types in Fig.~\protect\ref{fig:6types}
 with a repulsive two-body interaction with a strength that decreases exponentially. Note when the range of interactions is further than the barrier width, then no sectors are separable in any of the six cases. For zero range interactions, sectors $\mathcal{X}_B$ and $\mathcal{X}_D$ remain separable.}
\label{fig:types2int}%
\end{figure}

It is mathematically equivalent, and sometimes conceptually convenient, to treat a system with two particles in one dimension as though it were a system with one particle in two dimensions. When the barrier between the wells prevents tunneling, then the system is equivalent to one particle trapped four (not necessarily identical) two-dimensional wells.  Fig.~\ref{fig:types2} depicts sample two-dimensional potential energies $V(x_1) + V(x_2)$,
where $V(x_i)$ one of the six cases of single-particle double-wells. Fig.~\ref{fig:types2int}  depicts the same six potentials with the addition finite-range, repulsive interactions $g\exp(-\lambda|x_i - x_2|)$. The local and global configuration space symmetries can be classified, and these symmetries depend on the nature of the wells and the nature of the interactions. The cases of non-interacting,  finite-range interactions, zero-range interactions, and the unitary limit of contact interactions have different symmetries, and these are listed for the six cases in Tab.~\ref{tab:2types}. When there is tunneling, all the parity-symmetric and parity-asymmetric cases collapse and the analysis is much simpler, as shown in Tab.~\ref{tab:2typestun}.

Most of the configuration space analysis summarized in Tabs.~\ref{tab:2types} and \ref{tab:2typestun} can be inferred directly from Figs.~\ref{fig:types2} and \ref{fig:types2int}. The groups represent all the piecewise-linear transformations that map the configuration space onto itself. Despite the rich symmetry structures possible, the degeneracy of states for two particles in two wells generally cannot be explained by the irreps of the configuration space symmetry group. Like the one-particle case, the full kinematic symmetry group of the Hamiltonian is required.  Subsection~\ref{sect:twopartsep} describes the kinematic symmetry group for any separable system and subsections \ref{sect:twopartwell} and \ref{sect:twopartord} describe how well permutation symmetry and ordering permutation symmetry manifest for two particles.

\begin{table*}%
\begin{tabular}{c|c|c|c|c}
\hline
Type & No interactions & Finite-range interactions & Contact interactions & Unitary limit \\
\hline
I & $\mathrm{D}_1\times\mathrm{W}_2\times\mathrm{D}_1$ & $\mathrm{D}_1\times\mathrm{W}_2\times\mathrm{D}_1$ & $\mathrm{D}_1\times\mathrm{W}_2\times\mathrm{D}_1$ & $\mathrm{O}_2\times\mathrm{W}_2\times\mathrm{O}_2$ \\
II & $\mathrm{D}_4\times\mathrm{D}_1\wr\mathrm{W}_2\times\mathrm{D}_1$ & $\mathrm{D}_2\times\mathrm{W}_2\times\mathrm{D}_1$ & $\mathrm{D}_2\times\mathrm{D}_1\wr\mathrm{W}_2\times\mathrm{D}_1$ & $\mathrm{D}_1\wr\mathrm{O}_2\times\mathrm{D}_1\wr\mathrm{W}_2\times\mathrm{O}_2$ \\
III & $\mathrm{D}_4\times\mathrm{D}_2\wr\mathrm{W}_2\times\mathrm{D}_4$ & $\mathrm{D}_2\times\mathrm{W}_2\times\mathrm{D}_2$ & $\mathrm{D}_2\times\mathrm{D}_2\wr\mathrm{W}_2\times\mathrm{D}_1$ & $\mathrm{D}_1\wr\mathrm{O}_2\times\mathrm{D}_2\wr\mathrm{W}_2\times\mathrm{D}_1\wr\mathrm{O}_2$ \\
IV & $\mathrm{D}_1\wr\mathrm{W}_4$ & $\mathrm{D}_1\wr\mathrm{W}_2\times\mathrm{W}_2$ & $\mathrm{D}_1\wr\mathrm{W}_2\times\mathrm{D}_1\wr\mathrm{W}_2$ & $\mathrm{O}_2\wr\mathrm{W}_2\times\mathrm{D}_1\wr\mathrm{W}_2$ \\
V & $\mathrm{D}_1\wr\mathrm{W}_4$ & $\mathrm{D}_1\wr\mathrm{W}_2\times\mathrm{D}_1\wr\mathrm{W}_2$ & $\mathrm{D}_1\wr\mathrm{W}_2\times\mathrm{D}_1\wr\mathrm{W}_2$ & $\mathrm{O}_2\wr\mathrm{W}_2\times\mathrm{D}_1\wr\mathrm{W}_2$ \\
VI & $\mathrm{D}_4\wr\mathrm{W}_4$ & $\mathrm{D}_2\wr\mathrm{W}_2\times\mathrm{D}_1\wr\mathrm{W}_2$ & $\mathrm{D}_2\wr\mathrm{W}_2\times\mathrm{D}_4\wr\mathrm{W}_2$ & $\mathrm{D}_1\wr\mathrm{O}_2\wr\mathrm{W}_2\times\mathrm{D}_4\wr\mathrm{W}_2$\\
\hline
\end{tabular}
\caption{For each of the six types of \emph{impenetrable} double-wells depicted in Figs.~\protect\ref{fig:types2} and \protect\ref{fig:types2int},
 the configuration space symmetries for two particles that are non-interacting, interacting via an interaction with range wider than the barrier domain, interacting via a contact interaction, and the unitary limit of the contact interaction. The group $\mathrm{D}_j$ is the two-dimensional dihedral point group with $j$ reflections and $j-1$ rotations. The group $\mathrm{W}_j$ is the well permutation group and is isomorphic to the symmetric group $\mathrm{S}_j$. The ordering permutation group $\mathrm{O}_j$ is also isomorphic to the symmetric group $\mathrm{S}_j$. In these expressions, the wreath product is taken before the direct product in the order of operations, and the wreath product is associative so $(\mathrm{D}_1\wr\mathrm{O}_2)\wr\mathrm{W}_2 = \mathrm{D}_1\wr(\mathrm{O}_2\wr\mathrm{W}_2)$.}
\label{tab:2types}
\end{table*}

\begin{table}%
\begin{tabular}{c|c|c|c}
\hline
Type & No interactions & Interactions & Unitary limit \\
\hline
I-IV & $\mathrm{P}_2\sim\mathrm{D}_1$ & $\mathrm{P}_2\sim\mathrm{D}_1$ & $\mathrm{O}_2\sim\mathrm{P}_2\sim\mathrm{D}_1$  \\
V-VI & $\mathrm{O}(1)\wr\mathrm{P}_2\sim\mathrm{D}_4$ & $\mathrm{O}(1)\times\mathrm{P}_2\sim\mathrm{D}_2$ & $\mathrm{D}_1\wr\mathrm{O}_2$  \\
\hline
\end{tabular}
\caption{This table lists the configuration space symmetries for each of the six types of double-wells depicted in Figs.~\protect\ref{fig:types2} and \protect\ref{fig:types2int}
 when there is tunneling for two particles that are non-interacting, interacting, and interacting via a contact interaction at the unitary limit. The group $\mathrm{P}_j$ is the particle permutation group and is isomorphic to the symmetric group $\mathrm{S}_j$. See the caption of Tab.~\protect\ref{tab:2types} for additional notation.
}
\label{tab:2typestun}
\end{table}

\subsection{Symmetry of Separability}\label{sect:twopartsep}

First, consider two non-interacting particles in a single one-dimensional well with one-particle energy spectrum $\sigma_1 = \{\epsilon_0,\epsilon_1,\epsilon_2,\ldots\}$. The two-particle energy spectrum $\Sigma$ is just the sum of two copies of $\sigma_1$, and each energy level is either singly-degenerate like the ground state with energy $2 \epsilon_0$ or doubly-degenerate like the first excited state $\epsilon_0 + \epsilon_1$. The group that describes this kinematic symmetry  is $\mathrm{T}\wr\mathrm{P}_2$. This group is isomorphic to the one-particle, double-well kinematic group $\mathrm{T}_a\wr\mathrm{W}_2$ for cases IV and V, but the correspondence between energy levels and irreps is not the same. The difference is that $\mathrm{T}\wr\mathrm{P}_2$ is the symmetry of a system with two identical \emph{separable} degrees of freedom. The Hamiltonian can be written in terms of an operator like
\begin{equation}
\hat{H} = \hat{h}_1 \otimes \hat{\mathbb{I}} + \hat{\mathbb{I}} \otimes \hat{h}_2
\end{equation}
acting on a tensor product structure like
\begin{equation}
L^2(\mathcal{X}_1) \otimes L^2(\mathcal{X}_2)\sim L^2(\mathbb{R}^2),
\end{equation}
where $\mathcal{X}_i$ is the one-dimensional configuration space for each particle.
On the other hand, the one-particle, identical double-well kinematic group $\mathrm{T}_a\wr\mathrm{W}_2$ is the symmetry for a system with a single degree of freedom segmented into disjoint intervals. The Hilbert space for that problem is 
\begin{equation}\label{eq:plush}
L^2(\mathcal{X}_a) \oplus L^2(\mathcal{X}_b)\sim \mathbb{C}^2 \otimes L^2(\mathcal{X}_a).
\end{equation}
So, although the groups are isomorphic and have the same irreps, for $\mathrm{T}_a\wr\mathrm{W}_2$ not all of those irreps occur in the reduction of the Hilbert space (\ref{eq:plush}).

For non-interacting identical two particle systems, the kinematic group always has $\mathrm{T}\wr\mathrm{P}_2$ as a subgroup~\cite{Harshman2016_I}. Its subgroup $\mathrm{T}_1\times \mathrm{T}_2 \subset \mathrm{T}\wr\mathrm{P}_2$ is the product of two commuting continuous symmetry transformations in phase space. Since there are two degrees of freedom, in this case  separability symmetry is enough to guarantee Liouville integrability for two non-interacting particles.

Interactions break the symmetry of separability but preserve the subgroup of total time translations and particle exchange $\mathrm{T} \times \mathrm{P}_2$ for cases I-IV. Total parity is additionally preserved for cases V and VI, giving $\mathrm{T} \times \mathrm{O}(1) \times \mathrm{P}_2$. This is not enough symmetry to preserve integrability for general traps since there is only one continuous symmetry transformation. Additionally, these groups only have one-dimensional representations, so interactions split the degeneracies created by the symmetry of separability.

\subsection{Well permutation symmetry}\label{sect:twopartwell}

When there is no tunneling, the two particle configuration space $\mathcal{X}^2$ is split into four, decoupled sectors. Starting from the upper right corner in any subfigure of Fig.~\ref{fig:types2} and moving counterclockwise,  label these sectors $A$ (both particles in well $b$), $B$ (particle one in well $b$, particle two in well $a$), $C$ (both particles in well $a$) and $D$ (particle one in well $a$, particle two in well $b$). 
Piecewise linear transformations of configuration space $\mathcal{X}^2\sim\mathbb{R}^2$ that shuffle similar sectors and commute with the Hamiltonian are local symmetries.

For all six cases of double wells, sectors $B$ and $D$ are mirror images whether there are interactions or not. These are the sectors corresponding to one particle in each well, and their similarity is a consequence of particle permutation symmetry. However, global particle permutation symmetry does not just exchange sectors $B$ and $D$; it also reflects sectors $A$ and $C$ about the line $x_1=x_2$. However, when the double-wells are impenetrable, the operator $w_{BD}$ that exchanges wells $B$ and $D$ can be (in principle) implemented as a local symmetry independent of the operators $\sigma_A$ and $\sigma_C$ that reflect sectors $A$ and $C$ along the line $x_1 = x_2$. The global action of the particle exchange operator $p_{12}$ can be decomposed into a product of three operators:
\[
p_{12} = \sigma_A \sigma_C w_{BD}.
\]
Well permutations like $w_{BD}$ and `well-local' operators like $\sigma_A$ do not need to be physically feasible. The physical protocol that $\sigma_A$ represents ``exchange particle 1 and 2 if they are both in the right well, otherwise leave them alone'' does not need to have an active realization. Well permutations are still valid symmetries even if they are only passive transformations of coordinate systems.

To explain the degeneracies in the non-tunneling limit, the configuration space symmetry must be combined with independent time translations for each well. Without interactions, the Hamiltonian restricted to each of the four sectors is still separable. As an example, consider, case I with no tunneling and no interactions. The total kinematic symmetry can be written as
\begin{equation}
(\mathrm{T}_{a} \wr \mathrm{P}_2) \times ((\mathrm{T}_a \times \mathrm{T}_b) \wr \mathrm{W}_2) \times (\mathrm{T}_{b} \wr \mathrm{P}_2).
\end{equation}
The first factor $\mathrm{T}_a \wr \mathrm{P}_2$ and third factor $\mathrm{T}_b \wr \mathrm{P}_2$ are the symmetry of separability combined with particle permutation for the cases when both particles in the same wells $A$ and $C$. The second factor is the symmetry of separability combined with well permutation symmetry for the two cases of one particle in each well $B$ and $D$. Assuming no accidental degeneracies, irreps for the first and third factor subgroup are one- and two-dimensional, and the relevant irreps for the second factor subgroup are two-dimensional.

When interactions are turned on for case I, separability is broken and the kinematic symmetry that remains is
\begin{equation}
(\mathrm{T}_A \times \mathrm{P}_2) \times (\mathrm{T}_{BD} \wr \mathrm{W}_2) \times (\mathrm{T}_C \times \mathrm{P}_2).
\end{equation}
Now the first and third factor subgroups only have one-dimensional irreps. The well permutation symmetry for sectors $B$ and $D$ continues to hold, and so there are still irreps corresponding two-fold degenerate energy levels. Note that if the interaction is zero-range (or of a range shorter than the barrier width) then the sectors $B$ and $D$ retain their separability symmetry. This distinction does not change the symmetry for case I, but it becomes relevant for cases II-VI, and it is necessary to explain the degeneracy pattern found for Hamiltonian (\ref{eq:habg}) in Table~\ref{tab:deg} with $a\neq 0$.

The other cases that are important for subsequent examples with Hamiltonian (\ref{eq:habg}) are cases V and VI. For these cases, all four two-particle sectors are equivalent when there are no interactions, so the configuration space symmetry has a larger well permutation subgroup $W_4\sim\mathrm{S}_4$. Additionally, each well in case V has $\mathrm{D}_1\sim\mathrm{O}(1)$ symmetry (one reflection) and each well in case VI has $\mathrm{O}(1)\wr\mathrm{P}_2 \sim \mathrm{D}_4$
symmetry (four reflections, three rotations). So without interactions, the configuration space symmetries are $\mathrm{D}_1\wr\mathrm{W}_4$ for case V and  $\mathrm{D}_4\wr\mathrm{W}_4$ for case VI. Each well is separable, and therefore by including the kinematic symmetries this is extended to $(\mathrm{T}\wr\mathrm{P}_2)\wr\mathrm{W}_4$ for case V and  $((\mathrm{T}\times\mathrm{O}(1))\wr\mathrm{P}_2)\wr\mathrm{W}_4$ for case VI. Both these groups have four-fold and eight-fold degenerate irreps that correspond to two-particle, two-well infinite barrier energy levels.

Adding interactions distinguishes wells $A$ and $C$ from wells $B$ and $D$ and breaks separability. For finite range interactions, the remaining kinematic symmetries are
\begin{eqnarray}
(\mathrm{T}_{AC}\times\mathrm{P}_2)\wr\mathrm{W}_2 &\times& (\mathrm{T}_{BD}\times\mathrm{D}_1)\wr\mathrm{W}_2\\
(\mathrm{T}_{AC}\times\mathrm{D}_1\times\mathrm{P}_2)\wr\mathrm{W}_2 &\times& (\mathrm{T}_{BD}\times\mathrm{D}_1)\wr\mathrm{W}_2
\end{eqnarray}
for cases V and VI, respectively. Both factor subgroups have two-dimensional irreps that correspond to physical states. For contact interactions, the remaining kinematic symmetries are
\begin{eqnarray}
(\mathrm{T}_{AC}\times\mathrm{P}_2)\wr\mathrm{W}_2 &\times& (\mathrm{T}_{BD} \wr\mathrm{D}_1)\wr\mathrm{W}_2\\
(\mathrm{T}_{AC}\times\mathrm{D}_1\times\mathrm{P}_2)\wr\mathrm{W}_2 &\times& (\mathrm{T}_{BD} \times \mathrm{D}_1\wr\mathrm{P}_2\wr\mathrm{W}_2.
\end{eqnarray}
The second subgroup factor has two-dimensional irreps and four-dimensional irreps, and the four-dimensional irrep has the same energy as the two-dimensional irrep of the first subgroup factor. Therefore, there are six-fold degenerate energy levels in addition to two-fold degenerate levels. See Section~\ref{sect:habg} for examples with Hamiltonian (\ref{eq:habg}).

\subsection{Ordering permutation symmetry}\label{sect:twopartord}

The final two-particle, two-well symmetry to be discussed here occurs in the unitary limit of the contact interaction, or more generally when strong, finite-range repulsive interactions effectively split configuration space $\mathcal{X}^2$ into two identical sections, $\mathcal{X}_{I}$ for $x_1 < x_2$ and $\mathcal{X}_{I\! I}$ for $x_2 < x_1$. See Fig.~\ref{fig:quartic} for a depiction. Then each of these sectors can again be thought of as a one-particle, two-dimensional wells and piecewise linear transformations of $\mathcal{X}^2$ that commute with the Hamiltonian are configuration space symmetries. Denote the group by $\mathrm{O}_2$. One might think this would be equivalent to particle exchange, and for asymmetric wells and asymmetric double wells like cases I-IV, one would be right. However, for symmetric traps (including double wells like cases V and VI), each section also allows an \emph{independent} reflection. The total configuration space symmetry group is therefore at least as large as $\mathrm{D}_1\wr\mathrm{O}_2$.

When there is a global parity-symmetric double-well with no transmission like cases V and VI, then there are two wells $A$ and $C$ that have ordering permutation symmetry. Therefore case V has total configuration space symmetry of
\begin{equation}\label{eq:cfVu}
(\mathrm{O}_2\wr\mathrm{W}_2) \times (\mathrm{D}_1\wr\mathrm{W}_2),
\end{equation}
 where the first factor subgroup is for sectors $A$ and $C$ and the second is for sectors $B$ and $D$. For case VI the configuration space symmetry is
\begin{equation}\label{eq:cfVIu}
(\mathrm{D}_1\wr\mathrm{O}_2\wr\mathrm{W}_2) \times (\mathrm{D}_4\wr\mathrm{W}_2).
\end{equation}
The final step is to include the kinematic symmetries, which enlarge (\ref{eq:cfVu}) and (\ref{eq:cfVIu}) into
\begin{equation}
(\mathrm{T}_{AC}\wr\mathrm{O}_2\wr\mathrm{W}_2) \times (\mathrm{T}_{BD}\wr\mathrm{D}_1\wr\mathrm{W}_2)
\end{equation}
and
\begin{equation}
((\mathrm{T}_{AC}\times\mathrm{D}_1)\wr\mathrm{O}_2\wr\mathrm{W}_2) \times ((\mathrm{T}_{BD}\times\mathrm{D}_1)\wr\mathrm{P}_2\wr\mathrm{W}_2).
\end{equation}
For both cases the physical irreps for the first factors have dimension four and for the second factor dimension two or four. Because the spectrum of sectors A and C shares energy levels with the spectrum of sectors B and D, the energy levels are either two-fold or eight-fold degenerate. 

Note that for more than two particles, ordering permutation symmetry has a richer structure~\cite{Harshman2016_II, Harshman2017} because the number of orders $N!$ is greater than the number of particles $N$.

\section{Split well Hamiltonian}\label{sect:habg}

This section applies kinematic symmetries to analyze the model Hamiltonian (\ref{eq:habg}). Physically, one can imagine this Hamiltonian is a good model for a highly-elongated, effectively one-dimensional trap that has been severed in two pieces by a sheet of repulsively-tuned light. For $a\neq 0$ the Hamiltonian has the well pattern of case I and for $a = 0$ it is case V generally and case VI for the infinite square well.

Two observations are worth emphasizing:
\begin{itemize}
\item The Hamiltonian with no interactions but any barrier strength $H^\tau_0$ is integrable in the Liouvillian sense. The Hamiltonian is separable into two one-particle systems and the energy of each particle is an independent integrable of the motion.
\item In the unitary limit of contact interactions $\gamma\to\infty$, the Hamiltonian $H^\tau_\infty$ is again integrable, now in the Bethe-ansatz sense. The infinite barrier along $x_1=x_2$ provides diffractionless scattering and solutions in each sector are constructed by superpositions of separable solutions. This is the essence of the famous observation of fermionization of trapped hard-core bosons by Girardeau~\cite{girardeau_relationship_1960}.
\end{itemize}
For any other value of interaction strength, the Hamiltonian generally is not integrable in either sense. The notable exceptions of the  infinite square well and harmonic trap are treated in the next section. For other trap shapes, the interacting region of model space can be interrogated by perturbation expansions, exact diagonalization, or variational methods using eigenstates of the non-interacting limit $H^\tau_0$. One can also do perturbation expansions (or other methods) from the unitary limit $H^\tau_\infty$, but some caution is required because wave functions for the finite energy eigenstates of $H^\tau_\infty$ necessarily have nodes along the line $x_1=x_2$. No superposition of $H^\tau_\infty$ eigenstates will ever have a non-zero wave functions along the line $x_1=x_2$. Equivalently, the operator $H^\tau_\infty$ is not self-adjoint on $L^2(\mathbb{R}^2)$, but only on the subdomain
\[
L^2(\mathbb{R}^2/\{{\bf x}|x_1 = x_2\}) = L^2(\mathcal{X}_I) \oplus L^2(\mathcal{X}_{I\! I}) \sim \mathbb{C}^2 \otimes L^2(\mathcal{X}_I).
\]
As a consequence, the first order perturbed state and second order perturbed energy require renormalization~\cite{sen_perturbation_1999, belloni_infinite_2014, gharashi_one-dimensional_2015}. Note that a similar difficulty would also exist when using perturbation theory to extrapolate from the solutions of $H^\infty_\gamma$ (which is solvable for certain traps shapes) to $H^\tau_\gamma$. Another idea is to use variational methods to interpolate between the ground state of $H^\tau_0$ and the ground state of $H^\tau_\infty$ (or the lowest state in a symmetrized sector, see below) to approximate the ground state as a function of $\gamma$~\cite{barfknecht_correlation_2016}.

\subsection{Asymmetric case $a\neq 0$}

\begin{figure}
\includegraphics[width=\columnwidth]{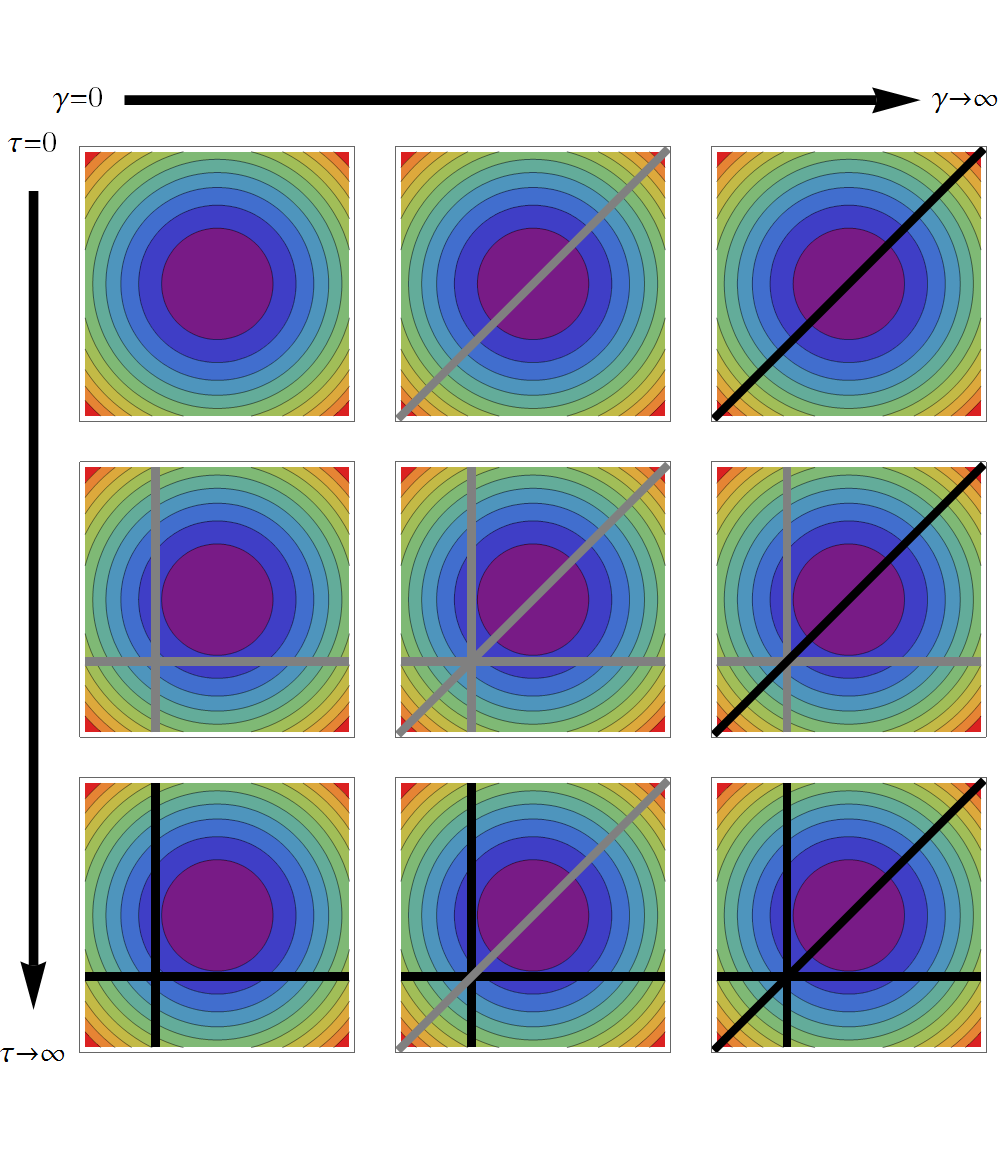}%
\caption{Potential energy of Hamiltonian (\protect\ref{eq:habg})  
for a harmonic trap with $a\neq 0$. Same ordering of subfigures as in Fig.~\protect\ref{fig:quartic}.
 Note that the left column and right column are integrable and are analytically solvable to arbitrary precision in terms of transcendental equations with parabolic cylindrical functions. \label{fig:offharm}}
\end{figure}

For the asymmetric case, the only symmetry valid for all $\tau$ and $\gamma$ is particle permutation $\mathrm{P}_2$. This means the Hilbert space can be reduced into permutation symmetric and antisymmetric sectors:
\begin{equation}\label{eq:redasym}
\HS = \HS^{[2]} \oplus \HS^{[1^2]},
\end{equation}
where we use the notation $[2]$ for symmetric and $[1^2]$ for antisymmetric.
There are no matrix elements of $H^\tau_\gamma$ (or any operator that has $\mathrm{P}_2$ symmetry) between vectors in different sectors. Note that in Fig.~\ref{fig:offharm}, which schematically depicts the potentials for a harmonic trap split by an off-center barrier, all subfigures have reflection symmetry across the line $x_1=x_2$. When $\tau=0$, parity is restored and so the potentials for the Hamiltonians depicted in the top line of Fig.~\ref{fig:offharm} all have at least $\mathrm{D}_2$ symmetry, which includes two reflections and one $\pi$ rotation. The left column depicts the limit $\gamma=0$ where the symmetry of separability applies; the bottom row depicts the limit $\tau\to\infty$ where well permutation symmetry applies; and the right column depicts the limit $\gamma\to\infty$ where ordering permutation symmetry applies. When there are additional symmetries, there are additional reductions of the Hilbert space into `smaller' sectors.

For an arbitrary $a \neq 0$, any one-particle energy degeneracies between the left and right sides of the barrier are accidental. For example, such a degeneracy would occur if the zero-range barrier was raised exactly at the nodal point of single-particle energy eigenfunction. Neglecting this kind of idealization, then the degeneracies of $H^\tau_\gamma$ for $a \neq 0$ should follow the analysis for case I of the previous section and summarized in the first column of Tab.~\ref{tab:deg}.

Denote the energy spectrum and energy eigenstates for the single-particle Hamiltonian $\hat{h}_1$ by $\sigma_1=\{\epsilon_0,\epsilon_1,\ldots\}$ and $\{\kt{0},\kt{1},\ldots\}$. The familiar solutions for $H^0_0$ that transform irreducibly under $\mathrm{P}_2$ are algebraically constructed from the single particle solutions using superpositions of tensor product states $\kt{n_1 n_2}=\kt{n_1}\otimes\kt{n_2}$:
\begin{subequations}\label{eq:00symm}
\begin{eqnarray}
\kt{(n_1n_2) +} &=& \left\{ \begin{array}{cl} \kt{n_1 n_2} & n_1 = n_2\\
\frac{1}{\sqrt{2}}\left(  \kt{n_1n_2} + \kt{n_2n_1} \right) & n_1 \neq n_2 \end{array} \right. \\
\kt{(n_1n_2) -} &=&  \frac{1}{\sqrt{2}}\left( \kt{n_1n_2} - \kt{n_2n_1} \right)\label{eq:00symm-}.
\end{eqnarray}
\end{subequations}
In this expression and in subsequent notations, without loss of generality the composition $(n_1n_2)$ of two quantum numbers is ordered so that $n_1 \leq n_2$. The spectrum of $H^0_0$ is the sumset (or Minkowski sum) $\Sigma_0^0 = \sigma_1 + \sigma_1$ with elements like $E_{n_1n_2}=\epsilon_{n_1}+\epsilon_{n_2}$. As expected from the kinematic symmetry of separability, energies $E_{n_1n_2}$ are one-fold or two-fold degenerate depending on whether $n_1$ equals $n_2$. Note that without more information about the set $\sigma_1$, only a partial order can be placed on $\Sigma_0^0$. For example, $E_{00} < E_{01} < E_{11}$, but without more information about the progression of energies in $\sigma_1$ we cannot know whether $E_{11}$ or $E_{02}$ is greater.

Adding a permeable barrier, the Hamiltonian $H^\tau_0$ is separable for any $\tau$ and denote the single-particle energy eigenstates of $h_1 + \tau\delta(x -a)$ by $\kt{n}^\tau$ with energy $\epsilon_n^\tau\in\sigma^\tau_1$ for $n$ a non-negative integer. Denote the eigenvectors of $H^\tau_0$ by $\kt{n_1n_2}^\tau = \kt{n_1}^\tau \otimes \kt{n_2}^\tau$; they can be permutation symmetrized as in (\ref{eq:00symm}) to form $\kt{(n_1n_2)^\tau +}\in\HS^{[2]}$ and $\kt{(n_1n_2)^\tau -}\in\HS^{[1^2]}$. In addition to the Hilbert space reduction (\ref{eq:redasym}), for the Hamiltonian $H^\tau_0$ and alternate reduction that exploits separability is
\begin{equation}
\HS^\tau_0 = \bigoplus_{(n_1n_2)} \HS^{(n_1n_2)^\tau_0},
\end{equation}
where the sum is over all compositions of two non-negative integers. Because the Hamiltonian $H^\tau_0$ is separable for any $\tau$, the quantum numbers $n_1$ and $n_2$ are conserved as $\tau$ is varied adiabatically, even though the state $\kt{n_1n_2}^\tau$ and the energy $E^\tau_{n_1n_2} = \epsilon_{n_1}^\tau + \epsilon_{n_2}^\tau$ change.

In contrast, when interactions are turned on, the symmetry of separability is broken and the two-fold degenerate levels of $\Sigma^\tau_0 = \sigma^\tau_1 + \sigma^\tau_1$ split into singly-degenerate levels. The states $\kt{(n_1n_2)^\tau -}$ have nodes along $x_1=x_2$ and so they do not feel the zero-range contact interaction, whereas the symmetric states $\kt{(n_1n_2)^\tau +}$ shift upwards in energy in a fashion that generally depends on the shape of the trap, strength of the barrier, and the quantum numbers in the composition.

The eigensolutions at the unitary limit $H^\tau_\infty$ can be algebraically constructed from the eigensolutions of $H^\tau_0$ by restricting the particle permutation antisymmetric states (\ref{eq:00symm-}) to the domains $\mathcal{X}_I$ and $\mathcal{X}_{I\!I}$, also called the `snippet' basis~\cite{deuretzbacher_exact_2008, fang_exact_2011, Harshman2016_I, Harshman2016_II, decamp_exact_2016}:
\begin{eqnarray}\label{eq:snip0i}
\psi^I_{(n_1n_2)^\tau} ({\bf x}) &=&  \left\{ \begin{array}{ll} \sqrt{2} \psi_{(n_1n_2)^\tau -} ({\bf x}) & {\bf x} \in \mathcal{X}_I \\ 0 & {\bf x} \in \mathcal{X}_{I\! I} \end{array} \right.\\
\psi^{I\! I}_{n_1n_2} ({\bf x}) &=&  \left\{ \begin{array}{ll} -\sqrt{2} \psi_{(n_1n_2)^\tau -} ({\bf x}) & {\bf x} \in \mathcal{X}_{I\! I} \\ 0 & {\bf x} \in \mathcal{X}_I \end{array} \right. ,
\end{eqnarray}
where
\[
\psi_{(n_1n_2)^\tau -} ({\bf x}) = \bk{{\bf x}}{(n_1n_2)^\tau -}\]
and
\[
\psi^I_{(n_1n_2)^\tau} ({\bf x}) = \bk{{\bf x}}{(n_1n_2)^\tau; I}.
\]
For every energy level of the non-interacting spectrum $\Sigma^\tau_0$, there is a two-fold degenerate level in the spectrum of $H^\tau_\infty$, denoted $\Sigma^\tau_\infty$. From the degenerate eigenvectors (\ref{eq:snip0i}), the simultaneous eigenvectors of energy and particle permutation can be constructed:
\begin{eqnarray}
\kt{(n_1n_2)^\tau_\infty; [2] } &=& \frac{1}{\sqrt{2}} \left( \kt{(n_1n_2)^\tau; I} + \kt{(n_1n_2)^\tau; I\! I}\right)\\
\kt{(n_1n_2)^\tau_\infty; [1^2] } &=& \frac{1}{\sqrt{2}} \left(\kt{(n_1n_2)^\tau; I} - \kt{(n_1n_2)^\tau; I\! I} \right) \nonumber\\
&\equiv & \kt{(n_1n_2)^\tau [1^2]} \nonumber .
\end{eqnarray}
This is an example of bosonic fermionization a la Girardeau for the unitary limit of the contact interaction~\cite{PhysRevA.92.033621}. These two wave functions have the same energy and configuration space density, although they have different momentum distributions. 

A final comment before moving to the symmetric case: just like $H^\tau_0$, the integrability of $H^\tau_\infty$ means that the quantum number composition $(n_1n_2)$ of the states $\kt{(n_1n_2)^\tau_\infty; [2]  }$ and $\kt{(n_1n_2)^\tau_\infty; [1^2] }$ do not change as $\tau$ is adiabatically varied, even though the states and energies do.

\subsection{Symmetric case $a = 0$}

When the barrier is erected in the middle of the symmetric trap, then parity and permutation symmetry are preserved for all values of $\tau$ and $\gamma$. This means that all states can be reduced into one of four irreps of $\mathrm{P}_2 \times \mathrm{O}(2)$: $[2]^+$, $[2]^-$, $[1^2]^+$ and $[1^2]^-$. In other words, the total Hilbert space for the system is broken into four sectors
\begin{equation}
\HS^\tau_\gamma = \HS^{[2]^+} \oplus \HS^{[2]^-} \oplus \HS^{[1^2]^+} \oplus \HS^{[1^2]^-}
\end{equation}
and there are no matrix elements of the Hamiltonian $H^\tau_\gamma$ for vectors in different sectors. The parity of $H^\tau_0$ eigenstates $\kt{(n_1n_2)^\tau +}$ and $\kt{(n_1n_2)^\tau -}$ are $(-1)^{n_1 + n_2}$. For the $H^\tau_\infty$ eigenstates, the permutation symmetric state $\kt{(n_1n_2)^\tau_\infty; [2] }$ has parity $ -(-1)^{n_1 + n_2}$, whereas for the antisymmetric state $\kt{(n_1n_2)^\tau_\infty; [1^2] }$ the parity has the normal form $ (-1)^{n_1 + n_2}$.
 
Another consequence of the centered barrier is that solutions for $H^\infty_0$ (which has well permutation symmetry) and $H^\infty_\infty$ (which has well and ordering permutation symmetry) can also be constructed from the eigensolutions of $H^0_0$. This occurs because the barrier is raised exactly where every negative parity single-particle eigenstate $\kt{n_i}$ of $\hat{h}_i$ has a node. The spectrum $\Sigma_0^\infty$  is the subset of $\Sigma_0^0$ derived from compositions of odd positive integers. Following the same argument as the algebraic construction of the unitary limit solutions of $H^\tau_\infty$, the odd-parity single particle sates can be used to form a snippet basis for $H^\infty_0$. Each composition $(n_1n_2)$ of two odd states leads to a four-fold ($n_1=n_2)$ or an eight-fold ($n_1 \neq n_2)$ degenerate energy level (assuming no additional symmetries or accidental degeneracies), in agreement with the analysis of the kinematic symmetries and degeneracies for cases V and VI. To see this, define the quadrants in configuration space ${\bf x} = (x_1,x_2) \in \mathcal{X}$:
\begin{eqnarray*}
\mathcal{X}_A = \{ {\bf x}| x_1>0, x_2>0 \}, \mathcal{X}_B = \{ {\bf x}| x_1<0, x_2>0 \},\\
\mathcal{X}_C = \{ {\bf x}| x_1<0, x_2<0 \}, \mathcal{X}_D = \{ {\bf x}| x_1>0, x_2<0 \}.
\end{eqnarray*}
Then for a composition $(nn)$ with $n$ odd there are four `snippet' basis vectors, one for each quadrant. The snippet basis vector $\psi^A_{(nn)}({\bf x}) = \bk{{\bf x}}{nn;\! A}$ for the $\mathcal{X}_A$ quadrant is defined using the position representation $\psi_{(nn)}({\bf x})= \bk{{\bf x}}{nn}$, i.e.\
\begin{equation}
\psi^A_{(nn)}({\bf x}) = \left\{ \begin{array}{ll} 2 \psi_{nn}({\bf x}) & {\bf x} \in \mathcal{X}_A \\ 0 & {\bf x} \notin \mathcal{X}_A \end{array} \right. . 
\end{equation}
The same definition holds for $\kt{nn;C}$. For $\kt{nn;B}$ and $\kt{nn;D}$, the phase convention
\begin{equation}
\psi^B_{(nn)}({\bf x}) = \left\{ \begin{array}{ll} -2 \psi_{nn}({\bf x}) & {\bf x} \in \mathcal{X}_B \\ 0 & {\bf x} \notin \mathcal{X}_B \end{array} \right.  
\end{equation}
is convenient because all quadrants have the same phase.

Similarly,  the following eight degenerate energy eigenvectors can be defined when $n_1 \neq n_2$ using restrictions to quadrants and the same phase convention:
\begin{eqnarray*}
\kt{n_1n_2;\!A},\kt{n_2n_1;\!A},\kt{n_1n_2;\!B},\kt{n_2n_1\!;B},\\
\kt{n_1n_2;\!C},\kt{n_2n_1;\!C},\kt{n_1n_2;\!D},\kt{n_2n_1;\!D}.
\end{eqnarray*}

Simultaneous eigenvectors of energy, particle exchange, and parity can be constructed from these snippet vectors. For compositions of a single odd quantum number, they are
\begin{subequations}
\begin{eqnarray}
\kt{(nn)_0^\infty;\! [2]^+ ;\! 1} &=& \frac{1}{\sqrt{2}} \left( \kt{nn ;\! A} + \kt{nn ;\! C} \right)\label{eq:nn1}\\
\kt{(nn)_0^\infty;\! [2]^+ ;\! 2} &=& \frac{1}{\sqrt{2}} \left( \kt{nn;\! B} + \kt{nn;\! D}\right)\label{eq:nn2}\\
\kt{(nn)_0^\infty;\! [2]^- } &=& \frac{1}{\sqrt{2}} \left( \kt{nn;\! A} - \kt{nn ;\!C}\right)\\
\kt{(nn)_0^\infty;\! [1^2]^- } &=& \frac{1}{\sqrt{2}} \left( \kt{nn;\! B} - \kt{nn ;\!D}\right).
\end{eqnarray}
\end{subequations}
Note that the two states (\ref{eq:nn1}) and (\ref{eq:nn2}) have the same energy, parity and exchange symmetry, so any linear combination of them is also a simultaneous eigenvector of the three symmetry operators. For compositions of two odd numbers $(n_1n_2)$ are
\begin{eqnarray*}
\kt{(n_1n_2)_0^\infty;\! [2]^+ ;\! 1} &=& \frac{1}{\sqrt{2}} \left( \kt{(n_1n_2) + ;\! A} + \kt{(n_1n_2) + ;\! C} \right)\\
\kt{(n_1n_2)_0^\infty;\! [2]^+ ;\! 2} &=& \frac{1}{\sqrt{2}} \left( \kt{(n_1n_2) + ;\! B} + \kt{(n_1n_2) + ;\! D} \right)\nonumber\\
\kt{(n_1n_2)_0^\infty;\! [2]^- ;\! 1} &=& \frac{1}{\sqrt{2}} \left( \kt{(n_1n_2) + ;\! A} - \kt{(n_1n_2) + ;\! C} \right)\nonumber\\
\kt{(n_1n_2)_0^\infty;\! [2]^- ;\! 2} &=& \frac{1}{\sqrt{2}} \left( \kt{(n_1n_2) - ;\! B} - \kt{(n_1n_2) - ;\! D} \right)\nonumber\\
\kt{(n_1n_2)_0^\infty;\! [1^2]^+ ;\! 1} &=& \frac{1}{\sqrt{2}} \left( \kt{(n_1n_2) - ;\! A} + \kt{(n_1n_2) - ;\! C} \right)\nonumber\\
\kt{(n_1n_2)_0^\infty;\! [1^2]^+ ;\! 2} &=& \frac{1}{\sqrt{2}} \left( \kt{(n_1n_2) - ;\! B} + \kt{(n_1n_2) - ;\! D} \right)\nonumber\\
\kt{(n_1n_2)_0^\infty;\! [1^2]^- ;\! 1} &=& \frac{1}{\sqrt{2}} \left( \kt{(n_1n_2) - ;\! A} - \kt{(n_1n_2) - ;\! C} \right)\nonumber\\
\kt{(n_1n_2)_0^\infty;\! [1^2]^- ;\! 2} &=& \frac{1}{\sqrt{2}} \left( \kt{(n_1n_2) + ;\! B} - \kt{(n_1n_2) + ;\! D} \right).\nonumber
\end{eqnarray*}
Every simultaneous eigenvector of $\mathrm{O}(1) \times \mathrm{P}_2$ is two-fold degenerate for these levels.  Note that six of these vectors have no support along the line $x_1=x_2$, either because they have no support in quadrants $\mathcal{X}_A$ and $\mathcal{X}_C$, or because they have nodes due to particle permutation antisymmetrization. These six states form an invariant subspace under variations of $\gamma$ because they do not feel the zero-range contact interaction. If a small finite range were included, this six-fold degeneracy would break into three two-fold degenerate levels.

Finally, these results are combined with the previous subsection to find the spectrum of $H^\infty_\infty$. Now there are six regions of configuration space, denoted $\mathcal{X}_{AI}$, $\mathcal{X}_{AI\!I}$, $\mathcal{X}_B$, $\mathcal{X}_{CI}$, $\mathcal{X}_{CI\!I}$, and $\mathcal{X}_D$. The spectrum of this Hamiltonian is the same as $H_0^\infty$, i.e.\ energies for every composition $(n_1n_2)$ of two odd positive integers. Compositions with $n_1=n_2$ are two-fold degenerate, and compositions $n_1 \neq n_2$ are eight-fold degenerate. Without presenting all the definitions but relying on the notation to carry the semantic load, the simultaneous eigenvectors for energy, particle permutation and parity for $n_1 = n_2 = n$ an odd positive integer are
\begin{eqnarray}
\kt{(nn)_\infty^\infty;\! [2]^+ } &=&  \frac{1}{\sqrt{2}} \left( \kt{nn;\! B} + \kt{nn;\! D}\right)\\
\kt{(nn)_\infty^\infty;\! [1^2]^- } &=& \frac{1}{\sqrt{2}} \left( \kt{nn;\! B} - \kt{nn ;\!D}\right),\nonumber
\end{eqnarray}
For $n_1 \neq n_2$ odd positive integers, the definitions are 
\begin{eqnarray*}
\kt{(n_1n_2)_\infty^\infty;\! [2]^+ ;\! 1} &=& \frac{1}{2} \left( \kt{(n_1n_2) - ;\! AI} + \kt{(n_1n_2) - ;\! AI\!I} \right.\\
&& {} \left. + \kt{(n_1n_2) - ;\! CI} + \kt{(n_1n_2) - ;\! CI\!I} \right)\nonumber\\
\kt{(n_1n_2)_\infty^\infty;\! [2]^+ ;\! 2} &=& \frac{1}{\sqrt{2}} \left( \kt{(n_1n_2)_+ ;\! B} + \kt{(n_1n_2)_+ ;\! D} \right)\nonumber\\
\kt{(n_1n_2)_\infty^\infty;\! [2]^- ;\! 1} &=&  \frac{1}{2} \left( \kt{(n_1n_2) - ;\! AI} + \kt{(n_1n_2) - ;\! AI\!I} \right.\nonumber\\
&& {} \left. - \kt{(n_1n_2) - ;\! CI} - \kt{(n_1n_2) - ;\! CI\!I} \right)\nonumber\\
\kt{(n_1n_2)_\infty^\infty;\! [2]^- ;\! 2} &=& \frac{1}{\sqrt{2}} \left( \kt{(n_1n_2) - ;\! B} - \kt{(n_1n_2) - ;\! D} \right)\nonumber\\
\kt{(n_1n_2)_\infty^\infty;\! [1^2]^+ ;\! 1} &=& \frac{1}{2} \left( \kt{(n_1n_2) - ;\! AI} - \kt{(n_1n_2) - ;\! AI\!I} \right.\nonumber\\
&& {} \left. - \kt{(n_1n_2) - ;\! CI} + \kt{(n_1n_2) - ;\! CI\!I} \right)\nonumber\\
\kt{(n_1n_2)_\infty^\infty;\! [1^2]^+ ;\! 2} &=& \frac{1}{\sqrt{2}} \left( \kt{(n_1n_2) - ;\! B} + \kt{(n_1n_2) - ;\! D} \right)\nonumber\\
\kt{(n_1n_2)_\infty^\infty;\! [1^2]^- ;\! 1} &=& \frac{1}{2} \left( \kt{(n_1n_2) - ;\! AI} - \kt{(n_1n_2) - ;\! AI\!I} \right.\nonumber\\
&& {} \left. + \kt{(n_1n_2) - ;\! CI} - \kt{(n_1n_2) - ;\! CI\!I} \right)\nonumber\\
\kt{(n_1n_2)_\infty^\infty;\! [1^2]^- ;\! 2} &=& \frac{1}{\sqrt{2}} \left( \kt{(n_1n_2) + ;\! B} - \kt{(n_1n_2) + ;\! D} \right).\nonumber
\end{eqnarray*}
If there were a finite range to the interaction, these eight levels would break into two two-fold degenerate levels for the states with two particles in the same well, and two two-fold degenerate levels for the particles in different wells.

\section{Examples of solvable models}\label{sect:solv}

The analysis of the limiting cases of $H_\gamma^\tau$ described above holds for any symmetric external trap potential $V(x)=V(-x)$. The summary insights are:
\begin{itemize}
\item For any $a$, the limiting cases of no interaction $H_0^\tau$ and infinite contact interactions $H_\infty^\tau$ are integrable for every $\tau$.
\item When $a \neq 0$, the energy spectrum (energies, degeneracies, and eigenstates) of $H^\tau_\infty$ can be determined from the spectrum of $H_\tau^0$ using algebraic methods and all states are classified by irreps of $\mathrm{P}_2$.
\item When $a = 0$, the energy spectrum of $H^0_\infty$, $H_0^\infty$ and $H_\infty^\infty$ can be determined from the spectrum of $H_0^0$ using algebraic methods and all states are classified by irreps of $\mathrm{O}(1) \times \mathrm{P}_2$.
\end{itemize}
To make these results more clear, and to show how the spectra map onto each other as parameters are varied, the next two subsections provide examples for two familiar traps with extra solvability (and therefore extra symmetry): infinite square well and harmonic trap. For these wells the Hamiltonians $H^0_\gamma$ and $H^\tau_\gamma$ are also solvable.

\subsection{Infinite square well}

For the infinite square well potential 
\begin{equation}
V(x) = \left\{ \begin{array}{ll} 0 & 0 < x < L \\ \infty & \mbox{else} \end{array}\right.
\end{equation}
(the $x$ origin has been shifted by $a/2$ for convenience of notation) has the familiar sinusoidal solutions
\begin{eqnarray}
\bk{{\bf x}}{n_1  n_2} &\equiv& \psi_{n_1n_2}({\bf x}) \\
&=& \frac{2}{L} \sin\left(\frac{(n_1+1) \pi x_1}{L}\right) \sin\left(\frac{(n_2+1) \pi x_2}{L}\right) \nonumber
\end{eqnarray}
with energies
\begin{eqnarray}
E_{n_1 n_2} &=& \frac{\hbar^2 \left((n_1+1)^2 + (n_2+1)^2\right) \pi^2}{2 m L^2} \\
&\equiv& \left((n_1+1)^2 + (n_2+1)^2\right) \epsilon_0. \nonumber
\end{eqnarray}
The stationary states provide the energy eigenbases of $H_0^0$, $H_0^\infty$, $H^0_\infty$ and $H_\infty^\infty$ with suitable superpositions and restrictions defined above. The properties of the lowest energy states for these four cases are summarized in Table \ref{tab:isw}.

\begin{table*}
\begin{tabular}{|ccccccccc|}
\hline
Energy 			& $2\epsilon_0$ & $5\epsilon_0$ & $8\epsilon_0$ & $10\epsilon_0$ & $13\epsilon_0$ & $16 \epsilon_0$ & $18\epsilon_0$ & $20\epsilon_0$ \\
\hline
Composition & $(00)$ & $(01)$ & $(11)$ & $(02)$ & $(12)$ & $(03)$ & $(22)$ & $(13)$ \\
 \hline \noalign{\smallskip}
\parbox[b]{2.8cm}{\center Symmetric \\  functions} & \includegraphics[width=40pt]{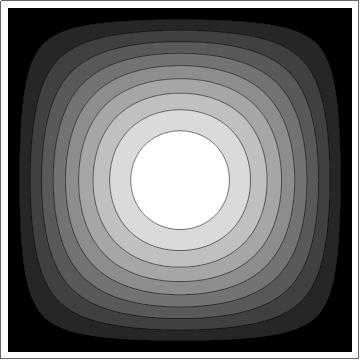}  & \includegraphics[width=40pt]{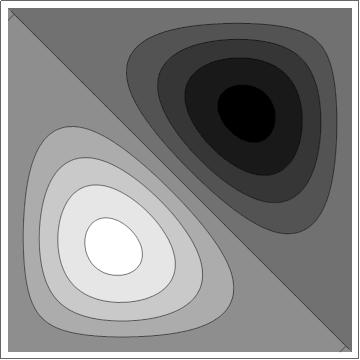} & \includegraphics[width=40pt]{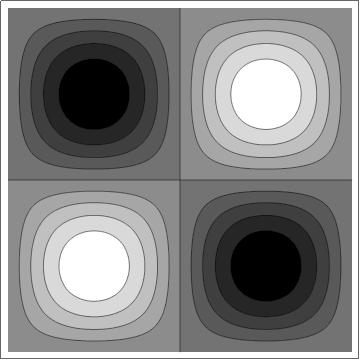} & \includegraphics[width=40pt]{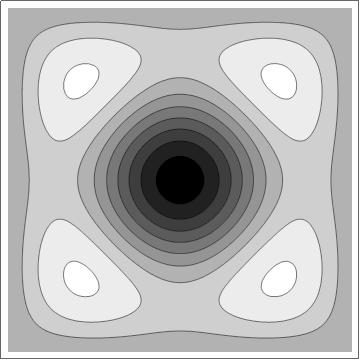} & \includegraphics[width=40pt]{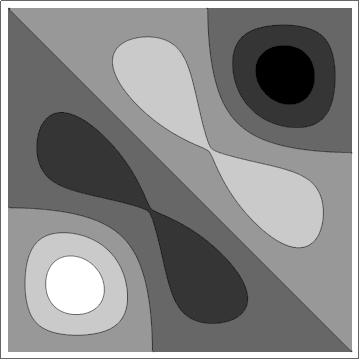} & \includegraphics[width=40pt]{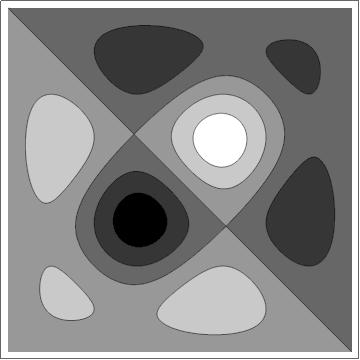} & \includegraphics[width=40pt]{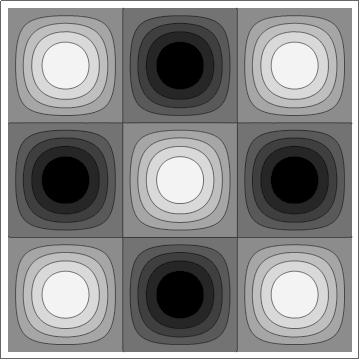} & \includegraphics[width=40pt]{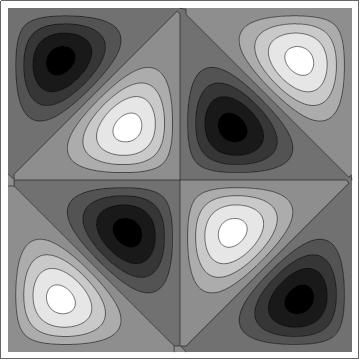} \\
\parbox[b]{2.8cm}{\center Antisymmetric \\  functions} &  & \includegraphics[width=40pt]{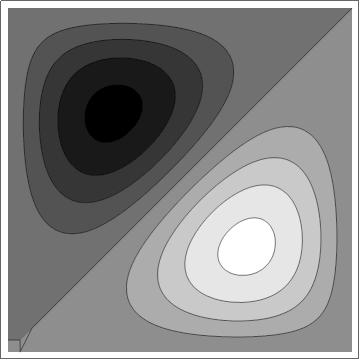} &  & \includegraphics[width=40pt]{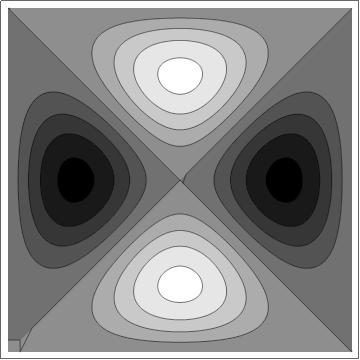} & \includegraphics[width=40pt]{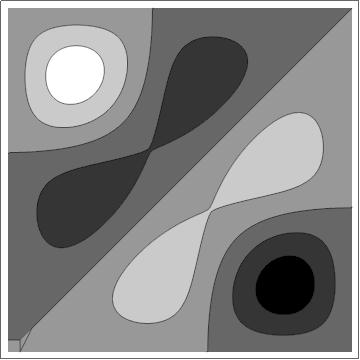} & \includegraphics[width=40pt]{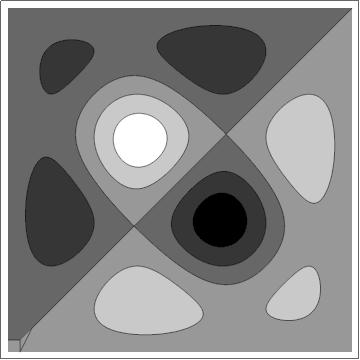} & & \includegraphics[width=40pt]{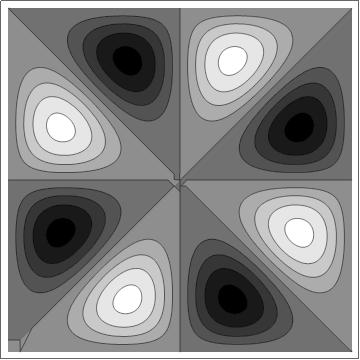} \\
\hline
$H_0^0$ & $\begin{array}{cc} 1 & 0 \\ 0 & 0 \end{array}$ & $\begin{array}{cc} 0 & 1 \\ 0 & 1 \end{array} $
& $\begin{array}{cc} 1 & 0 \\ 0 & 0 \end{array}$ & $\begin{array}{cc} 1 & 0 \\ 1 & 0 \end{array} $
& $\begin{array}{cc} 0 & 1 \\ 0 & 1 \end{array} $& $\begin{array}{cc} 0 & 1 \\ 0 & 1 \end{array}$ 
& $\begin{array}{cc} 1 & 0 \\ 0 & 0 \end{array}$ & $\begin{array}{cc} 1 & 0 \\ 1 & 0 \end{array}$ \\
\hline
$H_\infty^0$ & $\cdots$ & $\begin{array}{cc} 1 & 0 \\ 0 & 1 \end{array} $
 & $\cdots$ & $\begin{array}{cc} 0 & 1 \\ 1 & 0 \end{array}$
& $\begin{array}{cc} 1 & 0 \\ 0 & 1 \end{array} $& $\begin{array}{cc} 1 & 0 \\ 0 & 1 \end{array}$ 
& $\cdots$ & $\begin{array}{cc} 0 & 1 \\ 1 & 0 \end{array}$ \\
\hline
$H^\infty_0$ & $\cdots$ & $\cdots$
& $\begin{array}{cc} 2 & 1 \\ 0 & 1 \end{array}$ & $\cdots$
& $\cdots$ & $\cdots$ 
& $\cdots$ & $\begin{array}{cc} 2 & 2 \\ 2 & 2 \end{array}$ \\
\hline
$H^\infty_\infty$ & $\cdots$ & $\cdots$
& $\begin{array}{cc} 1 & 0 \\ 0 & 1 \end{array}$ & $\cdots$
& $\cdots$ & $\cdots$ 
& $\cdots$ & $\begin{array}{cc} 2 & 2 \\ 2 & 2 \end{array}$ \\
\hline
\end{tabular}
\caption{This table describes the lowest energy levels for $H_0^0$, $H_0^\infty$, $H^0_\infty$ and $H_\infty^\infty$ with the infinite square well potential. The composition tells what single-particle states make up the energy level. The inset figures for each energy are contour plots of $\bk{{\bf x}}{(n_1n_2)\! +}$ (for $n_1 = n_2$ or $n_1 \neq n_2$) and $\bk{{\bf x}}{(n_1n_2)\! -}$ (only for $n_1 \neq n_2$). The two-by-two arrays for each energy and each Hamiltonian show the degeneracy of the four types of irreducible representations of particle exchange symmetry and parity symmetry. The top row of each array is for $[2]^+$ and $[2]^-$, symmetric states with positive and negative parity. The bottom row is $[1^2]^+$ and $[1^2]^-$ for the antisymmetric states of both parities.}
\label{tab:isw}
\end{table*}

\begin{figure*}%
\includegraphics[width=\textwidth]{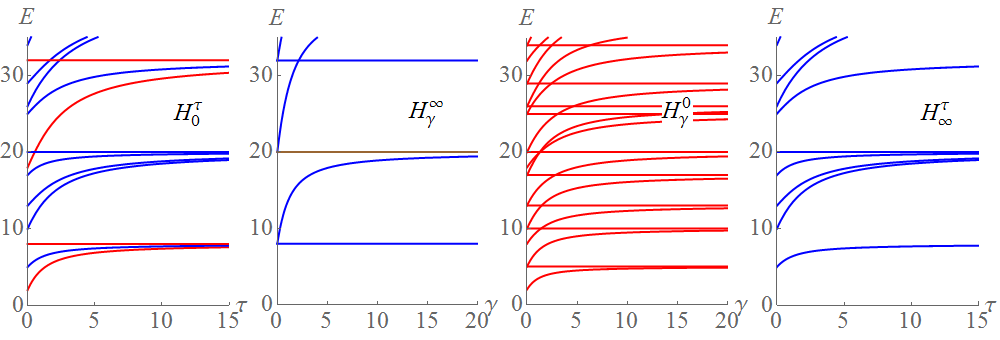}
\caption{Exact energy levels for $H_0^\tau$, $H_\gamma^\infty$, $H_\gamma^0$, and $H_\infty^\tau$ for two particles in the infinite square well potential. Energy is measured in units of $\epsilon_0$. Parameters $\tau$ and $\gamma$ are measured in units of $\epsilon_0 L$. Red lines indicate a non-degenerate energy level, blue lines are two-fold degenerate, and the brown line is six-fold degenerate.}%
\label{fig:isw}%
\end{figure*}

\begin{figure}%
\includegraphics[width=0.6\columnwidth]{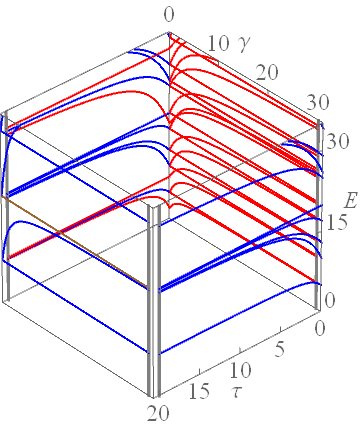}%
\caption{Same as %Fig.~\ref{fig:isw},
 but with four figures are combined into a three-dimensional representation that aligns with the edges in %Fig.~\ref{fig:quartic}.
 The gray vertical lines represent a break before the infinite limits. The splitting and merging of energy levels and the asymmetry between to two solvable paths from $H^0_0$ to $H^\infty_\infty$ can be seen.}%
\label{fig:isw2}%
\end{figure}

For the infinite square well potential, the Hamiltonians $H_0^\tau$, $H_\infty^\tau$, $H_\gamma^0$, and $H_\gamma^\infty$ are also solvable in terms of simple transcendental equations for any value of $\tau$ or $\gamma$.  As a result, the spectra at the four limiting cases $H_0^0$, $H_0^\infty$, $H^0_\infty$ and $H_\infty^\infty$ can be mapped to each other using explicit solutions. Although for any potential $H_0^\tau$ and $H_\infty^\tau$ are integrable, the integrability of $H_\gamma^0$ and $H_\gamma^\infty$ is special to homogeneous potentials and an example of a system where the Bethe ansatz works~\cite{cirone_bose-einstein_2001, oelkers_bethe_2006}. In Figs.~\ref{fig:isw} and \ref{fig:isw2}, the variation of the lowest energy levels are depicted. Summarizing observations about level mapping from these exact solutions:
\begin{itemize}
\item As $\tau$ increases from $0$ to $\infty$, the $H_0^0$ eigenstates with energy $E_{(n_1-1)(n_2-1)}$, $E_{n_1(n_2-1)}$, and $E_{(n_1-1)n_2}$ are mapped to $H_0^\infty$ eigenstates in the energy level $E_{n_1n_2}$ for $n_1$ and $n_2$ both odd. This holds for both four-fold and eight-fold degenerate levels of $H_0^\infty$.
\item As $\gamma$ increases from $0$ to $\infty$, the $H_0^0$ eigenstates $\kt{(n_1n_2)\!+}$ are mapped to $H_\infty^0$ eigenstates $\kt{n_1 (n_2+1)_\infty^0;\! [2]^\pi}$, where the parity is the same as the original state $\pi=(-1)^{n_1+n_2}$.
\item As $\tau$ increases from $0$ to $\infty$, the $H_\infty^0$ eigenstates follow the same rules for mapping to $H_\infty^\infty$ eigenstates as the $H_0^0$ eigenstates followed for mapping to $H_0^\infty$ eigenstates.
\item As $\gamma$ increases from $0$ to $\infty$, for $H_0^\infty$ energy levels with $n_1=n_2=n$, the eigenstates $\kt{(nn)_0^\infty; [2]^+ ;\! 1}$ and $\kt{(nn)_0^\infty;\! [2]^- }$ representing two particles in the same well and they are mapped to $H_\infty^\infty$ eigenstates $\kt{(n(n+2))_\infty^\infty;\! [2]^+ ;\! 1}$ and $\kt{(n(n+2))_\infty^\infty;\! [2]^- ;\! 1}$. The two states $\kt{(nn)_0^\infty;\! [2]^+ ;\! 2}$ and $\kt{(nn)_0^\infty;\! [1^2]^-}$ corresponding to particles in separate wells are invariant as $\gamma$ is changed. Similarly, for $H_0^\infty$ energy levels with $n_1 \neq n_2$, the two states $\kt{(n_1n_2)_0^\infty;\! [2]^+ ;\! 1}$ and $\kt{(n_1n_2)_0^\infty;\! [2]^- ;\! 1}$ shift to $\kt{(n_1(n_2+2))_\infty^\infty;\! [2]^+ ;\! 1}$ and $\kt{(n_1(n_2+2))_\infty^\infty; [2]^- ;\! 1}$, and the other six remain invariant.
\end{itemize}

What is the point of all this detailed analysis of level mapping? One consequence is that the order in which trap and interaction parameters are adiabatically tuned matters for state control. For example, start at $\tau = 0$ and $\gamma=0$ in the ground state $\kt{00}$. Slowly tune $\tau \to \infty$ and then $\gamma \to \infty$, and the state transforms as
\begin{eqnarray}
\kt{00} &\xrightarrow{\tau \to \infty}&  \frac{1}{\sqrt{2}}\left(\kt{(11)^\infty_0;\! [2]^+;\!1} +  \kt{(11)^\infty_0;\! [2]^+;\!2} \right)\nonumber\\
&\xrightarrow{\gamma \to \infty} &  \frac{1}{\sqrt{2}}\left(\kt{(13)^\infty_\infty;\! [2]^+;\!1} + \kt{(11)^\infty_\infty;\! [2]^+}\right)
\end{eqnarray}
which is a superposition of energy eigenstates. On the other hand, if the interactions are adiabatically ramped to the unitary limit and then the  tunneling is quenched, the state transforms as 
\begin{equation}
\kt{00} \xrightarrow{\gamma \to \infty} \kt{(01)_\infty^0;\! [2]^+} \xrightarrow{\tau \to \infty} \kt{(11)^\infty_\infty;\! [2]^+}
\end{equation}
into an energy eigenstate. If additionally, spin control were accessible and could change the effective symmetrization of a state, then it seems state control schemes that exploit these degeneracies are possible. However, an infinite square well with a zero-range barrier and interactions, tunable to extreme limits can only ever be an approximation. Then the question becomes, how sensitive is this kind of control to the vagaries of real experiments, where multiple assumptions may fail by a little or a lot?  That is a question for further work.

\subsection{Harmonic Trap}

For the harmonic oscillator trap, $\mathrm{U}(2)$ symmetry in phase space provides for additional systematic degeneracies, as well as renders 
 the four limiting cases $H_0^0$, $H_0^\infty$, $H^0_\infty$ and $H_\infty^\infty$ algebraically solvable. In fact there is enough additional symmetry so that those cases are superintegrable and solvable in both rectangular and polar coordinates. The $N$-th energy level of $H_0^0$ has $(N+1)$-fold degeneracy because of the equal energy level spacing, and then this degeneracy has consequences for the other three limiting cases; see Tab.\ \ref{tab:ho}.

\begin{table}
\begin{tabular}{|c|c|c|c|c|c|c|c|}
\hline
Energy 			& $\hbar\omega$ & $2\hbar\omega$ & $3\hbar\omega$ & $4\hbar\omega$ & $5\hbar\omega$ & $6\hbar\omega$ & $7\hbar\omega$   \\
\hline

$H_0^0$ & $\begin{array}{cc} 1 & 0 \\ 0 & 0 \end{array}$ & $\begin{array}{cc} 0 & 1 \\ 0 & 1 \end{array} $
& $\begin{array}{cc} 2 & 0 \\ 1 & 0 \end{array}$ 
& $\begin{array}{cc} 0 & 2 \\ 0 & 2 \end{array} $ & $\begin{array}{cc} 3 & 0 \\ 2 & 0 \end{array}$ & $\begin{array}{cc} 0 & 3 \\ 0 & 3 \end{array}$ & $\begin{array}{cc} 4 & 0 \\ 3 & 0 \end{array}$
 \\
\hline
$H_\infty^0$ & $\cdots$ & $\begin{array}{cc} 1 & 0 \\ 0 & 1 \end{array} $
& $\begin{array}{cc} 0 & 1 \\ 1 & 0 \end{array}$ 
& $\begin{array}{cc} 2 & 0 \\ 0 & 2 \end{array} $ & $\begin{array}{cc} 0 & 2 \\ 2 & 0 \end{array}$ & $\begin{array}{cc} 3 & 0 \\ 0 & 3 \end{array} $ & $\begin{array}{cc} 0 & 3 \\ 3 & 0 \end{array}$ 
 \\
 \hline
$H^\infty_0$ & $\cdots$ & $\cdots$
& $\begin{array}{cc} 2 & 1 \\ 0 & 1 \end{array}$ 
& $\cdots$ & $\begin{array}{cc} 2 & 2 \\ 2 & 2 \end{array}$ & $\cdots$ & $\begin{array}{cc} 4 & 3 \\ 2 & 3 \end{array}$
 \\
\hline
$H^\infty_\infty$ & $\cdots$ & $\cdots$
& $\begin{array}{cc} 1 & 0 \\ 0 & 1 \end{array}$ 
& $\cdots$ & $\begin{array}{cc} 2 & 2 \\ 2 & 2 \end{array}$ & $\cdots$ & $\begin{array}{cc} 3 & 2 \\ 2 & 3 \end{array}$
 \\
\hline
\end{tabular}
\caption{This table describes the lowest energy levels for $H_0^0$, $H_0^\infty$, $H^0_\infty$ and $H_\infty^\infty$ with the harmonic potential. The two-by-two arrays for each energy and each Hamiltonian show the degeneracy of the four types of irreducible representations of particle exchange symmetry and parity symmetry. The top row of each array is for $[2]^+$ and $[2]^-$, the symmetric states with positive and negative parity. The bottom row is $[1^2]^+$ and $[1^2]^-$ for the antisymmetric states of both parities.}
\label{tab:ho}
\end{table}

\begin{figure*}%
\includegraphics[width=\textwidth]{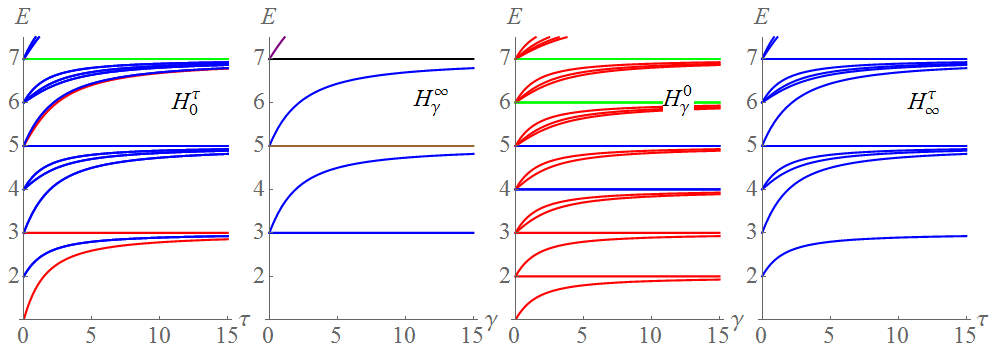}%
\caption{Exact energy levels for $H_0^\tau$, $H_\gamma^\infty$, $H_\gamma^0$, and $H_\infty^\tau$ for two particles in the harmonic potential. Energy is measured in units of $\hbar\omega$. Parameters and $\tau$ and $\gamma$ are measured in units of $\hbar\omega\sigma$, where $\sigma$ is the harmonic oscillator length scale. Red lines indicate a non-degenerate energy level and blue lines are two-fold degenerate. Green lines are three-fold degenerate, the purple line is four-fold degenerate, the brown line is six-fold degenerate, and the black line is eight-fold degenerate. }%
\label{fig:ho}%
\end{figure*}

As always, $H_0^\tau$ and $H_\infty^\tau$ are integrable. Further, $H_\gamma^0$ and $H_\gamma^\infty$ are also solvable~\cite{avakian_spectroscopy_1987, busch_two_1998, cirone_bose-einstein_2001} (but not algebraically solvable). The additional $\mathrm{U}(2)$ symmetry provided by the underlying isotropic harmonic trap allows separability in center-of-mass and relative coordinates for $H_\gamma^0$. The Hamiltonian $H_\gamma^\infty$ can be solved by defining an extension of the Hamiltonian in sectors $A$ and $C$ into sectors $B$ and $D$, and then forming suitable superpositions of separable solutions that solve the nodal boundary conditions at $x_1=0$ and $x_2=0$. The energy levels for all four limiting case Hamiltonians are depicted in Fig.\ \ref{fig:ho}. Numerical evidence from the exact solutions, confirmed by perturbation theory, shows that the additional degeneracies found for $H_0^0$, $H_0^\infty$, $H^0_\infty$ and $H_\infty^\infty$ with the harmonic oscillator trap do not exist for the non-algebraic solutions of $H_0^\tau$, $H_\infty^\tau$, $H_\gamma^0$, and $H_\gamma^\infty$. 

\section{More particles, more wells}\label{sect:multi}

The detailed symmetry analysis presented above may seem like overkill for finding the particular spectrum of a model with two-particles in one dimension since numerical methods converge speedily. Even for more complicated traps and interactions, many numerical approximation schemes can generate spectral results that are more accurate than the effective model and experimental control of any real experiment with ultracold atoms. What the symmetry analysis does provide is an analysis of what spectral features are universal for any trap and two-body interaction, and what is particular to specific traps, barriers and interactions. It also provides a mechanism for generating relations between models through symmetry breaking that can be used to track how states change under adiabatic tuning of model parameters. For systems with more degrees of freedom, where numerical methods are more difficult, symmetry methods may therefore provide a boost to model analysis.

The most important general results are:
\begin{itemize}
\item For non-interacting particles, no matter how many wells and how many particles, no matter the trap and the tunneling, the system is separable and Liouville integrable. It may of course be quite difficult to extract the spectrum for a strange trap shape, but in principle it is solvable with arbitrary accuracy.
\item At the unitary limit of contact interactions, the system is Bethe-ansatz integrable and solutions are constructed via the Girardeau mapping for any trap, barriers, and particle number.
\item Finally, for the infinite square well trap, for any number of particles and contact interactions of any strength, the system is integrable for no barriers and for infinite delta-barriers (or Heaviside step barriers). The Bethe ansatz solution for the finite interval works in each multiparticle square well.
\end{itemize}

\begin{figure}%
\includegraphics[width=0.33\columnwidth]{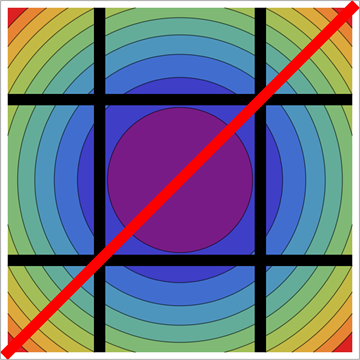}\includegraphics[width=0.33\columnwidth]{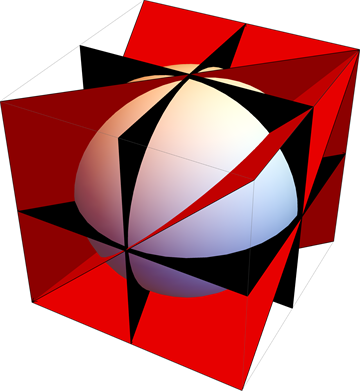}\includegraphics[width=0.33\columnwidth]{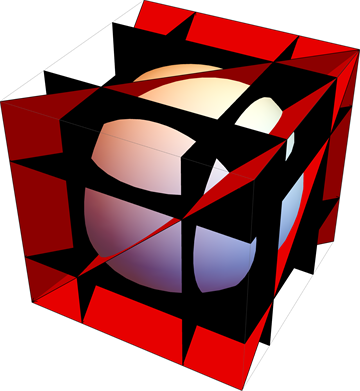}%
\caption{Potentials in configuration space for a harmonic trap split by thin barriers into well and sectioned by barriers (in black) and contact interactions (in red). The three cases are: (left) two particles in three wells with contact interactions; (center) three particles in two wells; (right) three particles in three wells. }%
\label{fig:part2well3}%
\end{figure}

As an example of the minimal case of an underfilled lattice with interacting particles, the first subfigure in Fig.~\ref{fig:part2well3} depicts the configuration space for two particles in a harmonic well that is split symmetrically by two barriers into three wells. In the limit of infinite barriers, there are nine two-particle sectors. The central sector, corresponding to both particles in the middle is non-degenerate, but the rest break into two quartets of exchangeable sectors. When there are no interactions, combining the separability with well permutation gives singly- or doubly-degenerate energy levels for both particles in the central well, and four- or eight-fold degenerate for one particle in middle well and one in edge well or both in edge wells. If the contact interaction is tuned to the unitary limit, the three sectors with particles in the same well split into subsectors, and the degeneracy is rearranged. There are two four-fold and two two-fold well permutation symmetries, with one of each being separable.

The simplest overfilling is realized by three particles in two wells, depicted in the middle subfigure in Fig.~\ref{fig:part2well3}. Without interactions, there is eight-fold well-permutation symmetry for a symmetrically-placed barrier in a symmetric trap. Contact interactions of finite strength break that symmetry into two-fold and six-fold exchangeable sectors, but the unitary limit increases it back up to two twelve-fold degenerate sectors. That means that the symmetry group has order $(12!)^2$, and that huge number make evident both the power and the limitations of exploiting this symmetry.

As a final comment on extensions, the three-particle, three-well case depicted in the right subfigure of Fig.~\ref{fig:part2well3} has  a symmetry group (in the case of no interactions or tunneling) that contains the Rubik's cube group as a subgroup~\footnote{This gives new meaning to solving the Rubik's cube ``the hard way'': design a quantum algorithm to exploit the symmetries of three distinguishable particles in three wells and then implement the protocol in a cold atom lab.}.

\section{Conclusions}

The group structure provided by impenetrable barriers, whether they are `real' barriers that partition traps into wells or `interaction' barriers that trap particles into specific orderings, is surprisingly rich, especially when parity is preserved. Even for two interacting particles in two wells, the possible symmetry structures are varied and complicated. Building up from the bottom, the order of finite symmetry group grows exponentially with the number of particles and impenetrable wells, even when the interactions are intermediate. This is promising, because the degrees of freedom are also growing rapidly, but it is also a challenge because working through the configuration space combinatorics is a non-trivial task and should only be undertaken if there is a meaningful pay-off. `Digitizing' the system by erecting impenetrable barriers seems to offer both promise and peril.

One avenue that looks productive is using these symmetries to induce degeneracies (and near-degeneracies) which can then be exploited for state control. One example for such an idea was presented in the section on the infinite square well, where it was shown that state evolution under adiabatic tuning of the interaction parameter and the tunneling parameter from zero to infinity depended on the order in which the tuning was performed. This is perhaps not surprising, but does open the possibility of using loops in parametrized `model space' to generate novel and useful quantum superpositions.

Another, more speculative idea is to see whether the symmetries of the few-body, few-well problems with unitary interactions and infinite barriers could be harnessed as an `analog quantum computer' for combinatorics problems. Akin to the boson sampling problem~\cite{v009a004} for calculating matrix permanents, perhaps there are combinatorics problems that (in the near future) would take fewer resources to embody in a few-body, few-well ultracold atomic system than to solve with traditional computers.

A final question is whether and how few-body, few-well models limit to the many-body, infinite lattice problem. Is it really practically to explore many-body physics from the bottom up? In particular, can the favorable growth of combinatoric symmetries for the strongly-interacting, weakly-tunneling be useful? This question and the previous ideas are worth further investigation.

\section*{Acknowledgments}

The author would like to thank the DC NASA Space Grant Foundation for support and to thank D.~Lockerby, F.F.~Bellotti, P.R.~Johnson, M.~Roberts, J.~Hirtenstein, M.~Olshanii, and M.A.~Garc\'ia-March for useful discussions. The author also thanks N.T.~Zinner and the Aarhus University Research Foundation for hosting his visit to Aarhus University where this paper was completed.

\end{document}